\documentclass[twocolumn,showpacs,nofootinbib,floats,amsmath,amssymb]{revtex4}
\usepackage{graphicx}
\usepackage{dcolumn}
\usepackage{bm}
\usepackage{amsmath}
\usepackage{float}
\usepackage{color}
\usepackage{amsfonts}
\usepackage{amssymb}
\usepackage[all]{xy}
\begin{document}
\title{Accelerated and decelerated expansion in a causal dissipative cosmology}

\author{Miguel Cruz}
\altaffiliation{miguelcruz02@uv.mx}

\affiliation{Facultad de F\'\i sica, Universidad Veracruzana 91000,
Xalapa, Veracruz, M\'exico}

\author{Norman Cruz}
\altaffiliation{norman.cruz@usach.cl}

\affiliation{Departamento de F\'{\i}sica, Universidad de Santiago,
Casilla 307, Santiago, Chile}

\author{Samuel Lepe}
\altaffiliation{samuel.lepe@pucv.cl}

\affiliation{Instituto de F\'{\i}sica, Facultad de Ciencias,
Pontificia Universidad Cat\'olica de Valpara\'\i so, Avenida Brasil
4950, Valpara\'\i so, Chile \\}

\date{\today}

\begin{abstract}
In this work we explore a new cosmological solution for an universe filled with one dissipative fluid, described by a barotropic EoS $p = \omega \rho$, in the framework of the full Israel-Stewart theory. The form of the bulk viscosity has been assumed of the form $\xi = \xi_{0}\rho^{1/2}$. The relaxation time is taken to be a function of the EoS, the bulk viscosity and the speed of bulk viscous perturbations, $c_{b}$. The solution presents an initial singularity, where the curvature scalar diverges as the scale factor goes to zero. Depending on the values for $\omega$, $\xi_{0}$, $c_{b}$ accelerated and decelerated cosmic expansion can be obtained. In the case of accelerated expansion, the viscosity drives the effective EoS to be of quintessence type, for the single fluid with positive pressure. Nevertheless, we show that only the solution with decelerated expansion satisfies the thermodynamics conditions $dS/dt > 0$ (growth of the entropy) and $d^{2}S/dt^{2} < 0$ (convexity condition). We show that an exact stiff matter EoS is not allowed in the framework of the full causal thermodynamic approach; and in the case of a EoS very close to the stiff matter regime, we found that dissipative effects becomes negligible so the entropy remains constant. Finally, we show numerically that the solution is stable under small perturbations.         
\end{abstract}


\maketitle

\section{Introduction}\label{intro}
Nowadays it is well established that cosmic evolution is in an accelerated phase \cite{riess, perlmutter, wmap}, however, a deeper understanding on the physical grounds of this acceleration is still lacking. As far as we know, after the inflation phase the universe was dominated by some components (as radiation) that eventually slow down the cosmic expansion, then the following question naturally arises: what is the unknown component that causes the acceleration observed?. Einstein himself addressed this problem by introducing the cosmological constant, however several problems are encountered in this approach even from the point of view of quantum field theory and recent results show some disparities with observations \cite{padilla, trodden}. One of the alternatives to reverse the slowing down effect of ordinary matter is the introduction of an exotic component called {\it dark energy}, whose main characteristic is that it has negative pressure. Although this exotic component solves the aforementioned problem, it faces other more serious such as the unknowing of its composition, the reason of its abundance in the universe and how it interacts with ordinary matter \cite{wang}. Over the years this has motivated the introduction of corrections to the General Relativity theory or consider some extensions of it in order to explore the cosmological consequences and although it is certain that some advance has been obtained, it is not at all conclusive, see for instance the Ref. \cite{clifton} for a review in modified gravity theories. In this proposal we will approach the subject of cosmic accelerated expansion by introducing viscosity effects through the causal Israel-Stewart (IS) theory \cite{Israel1976, Israel1979, viscous, maartens}. Such effects may play an important role in the evolution of relativistic fluids, see for example Ref. \cite{ref1} where was shown that a small amount of viscosity in cold dark matter fluid seems to alleviate a couple of discrepancies in the values obtained for some cosmological parameters when large scale structure (LSS) and Planck CMB data are used, in Ref. \cite{ref2} can be found a detailed analysis of viscosity effects in LSS formation and in the era of LIGO experiment the viscosity of dark matter and dark energy were constrained in \cite{ref3}.\\
 
In general, the most used theory to describe these irreversible processes is the Eckart theory, \cite{eckart} but has the problem of superluminal propagation velocities and some instabilities, however as a starting point has been very useful to understand the viscous effects in this framework \cite{observations}, such as the role of bulk viscosity for inflationary scenarios \cite{chitre} and some comparisons between both theories have been made, see Refs. \cite{comparing, competencia} for instance.\\

Within the viscous cosmological scenario it has gradually been shown that the above mentioned frameworks could be able to describe the late cosmic evolution and under the election of the appropriate Ansatz for the Hubble parameter a Big Rip scenario may be admissible, see the Refs. \cite{Cataldo2005, CruzLepe, CCL}, where the Eckart, IS and its non-linear extension theories were studied. This cosmic late time behavior was also found for an alternative viscous cosmological model \cite{scherrer}. An interesting review with a modern perspective on viscous cosmology can be found in \cite{saridakis}.\\
 
In this work we use an Ansatz for the Hubble parameter to solve the IS equation. This solution presents a singularity located in the remote past, so the universe has a finite age, and depending on the values of its parameters accelerated and decelerated expansion are possible. In the case of accelerated expansion,  the ordinary fluid behaves like quintessence,  wich has been invoked to describe the accelerated expansion in the cosmic evolution \cite{wang0}. In general, a quintessence fluid is described  by a scalar
field minimally coupled to gravity \cite{starobinsky}. A disadvantage of the scalar field is that there
is no direct evidence of its existence, one can say that this implies a decay process of the scalar field to the particles of the standard model, but formally this decay process also known as reheating must be introduced by hand in the context of standard inflation. In our approach a quintessential behavior can be obtained only by the introduction of dissipative effects at cosmological scales.\\

Since on physical grounds it is important to know how the accelerated and decelerated solutions behave at the thermodynamical level, we will explore the thermodynamics conditions $dS/dt > 0$ (growth
of the entropy) and $d^{2}S/dt^{2} < 0$ (convexity condition) in both cases.  As we shall see, the decelerated solution satisfy these conditions, so a universe filled with one fluid with positive pressure and viscosity can evolve, in the framework of the IS formalism, with a decelerated expansion, being this behavior thermodinamically consistent.\\
 
The outline of this paper is as follows: In Section \ref{sec:IStheory} we describe briefly the causal IS theory. In Section \ref{sec:new} we present a new solution for the IS theory, we discuss some of its characteristics and the three different types of cosmic evolution that can take place depending on the numerical value of our solution. In Section \ref{sec:thermo} we describe the thermodynamical behavior for our solution. In Section \ref{sec:elfinal} we give our final comments for this work. Finally, in an appendix we study the stability of this solution. $8\pi G=c=1$ units will be used in this work.

\section{Israel-Stewart theory}
\label{sec:IStheory}
In this section we will provide some highlights of the IS theory. The description of the universe is based on one main component given by a fluid and we will focus on two important features: i) obeys a barotropic EoS, $p = \omega \rho$, being $p$ the pressure and $\rho$ its density, where the parameter $\omega$ is constrained to the interval $[0,1)$, ii) along the cosmic evolution dissipative effects are present. For a flat FLRW universe the $tt$-component of the Einstein equations can be written as
\begin{equation}
3H^{2} = \rho, 
\label{eq:eq0}
\end{equation}
and the conservation equation for the dissipative fluid is given by
\begin{equation}
\dot{\rho} + 3H\left[ \left( 1+\omega \right) \rho +\Pi \right] =0,
\label{eq:eq18}
\end{equation}
where $\Pi$ is the viscous pressure which is governed by the following full transport equation \cite{Israel1979}
\begin{equation}
\tau \dot{\Pi}+\left( 1+\frac{1}{2}\tau \Delta \right) \Pi = -3 \xi (\rho) H, 
\label{eq:eq1}
\end{equation}
where the dot denotes derivatives with respect to the cosmic time. $\tau$ is the relaxation time for bulk viscous effects, in the limit $\tau = 0$ the theory is non-causal, $\xi(\rho)$ is the bulk
viscosity coefficient which depends on the energy density $\rho$, therefore $\xi \geq 0$, by $H$ we will denote the Hubble parameter, and $\Delta$ is defined by
\begin{equation}
\Delta := 3H+\frac{\dot{\tau}}{\tau }-\frac{\dot{\xi}}{\xi}-\frac{\dot{T}}{T}, 
\label{eq:eq2}
\end{equation}
where $T$ is the barotropic temperature, which takes the form $T = T_{0} \rho ^{\omega /\left(\omega +1\right)}$ once the Gibbs integrability condition is used, $T_{0}$ is a positive parameter. Additionally we have that \cite{maartens}
\begin{equation}
\frac{\xi}{\left(\rho +p\right)\tau} = c_{b}^{2},
\label{eq:relaxationtime}
\end{equation}
where $c_{b}$ is the speed of bulk viscous perturbations (non-adiabatic contribution to the speed of sound $V$), the dissipative speed of sound is given by $V^{2}= c_{s}^{2}+c_{b}^{2} \leq 1$, where the limit ensures causality and $c_{s}^{2}= (\partial p/\partial \rho)_{s}$ is the adiabatic contribution, for a barotropic EoS we simply have $c_{s}^{2}=\omega$ and we can write $c_{b}^{2}=\epsilon \left(1-\omega \right)$ with $0<\epsilon \leq 1$ (causality condition). We will adopt the standard assumption for the bulk viscosity coefficient, $\xi =\xi _{0}\rho^{s}$, being $\xi _{0}$ and $s$ positive constants \cite{barrow, viscous}. The relaxation time can be written as
\begin{equation}
\tau = \frac{\xi }{c_{b}^{2}\left( \rho +p\right) }=\frac{\xi }{%
c_{b}^{2}\rho \left( 1+\omega \right) }=\frac{\xi
_{0}}{\epsilon \left( 1-\omega ^{2}\right)}\rho ^{s-1}. 
\label{eq:eq3}
\end{equation}
Then, for $0\leq \omega <1$ and using Eqs. (\ref{eq:eq3}) and (\ref{eq:eq0}) we can write
\begin{equation}
\tau H=\frac{3^{s-1}\xi _{0}}{\epsilon \left( 1-\omega ^{2}\right) }%
H^{2\left( s-1/2\right)}.
\label{eq:eq7}
\end{equation}
As observed, for $s<1$, $s=1/2$ and $s<0$ we can have the limit $\tau \rightarrow 0$ as the Hubble parameter increases, in consequence we will have a rapid adjustment to the falling temperature of any particle species in order to maintain the thermal equilibrium with the cosmic fluid \cite{chimento}. From previous equations we can note a singular behavior for the relaxation time as we approach to the stiff matter case, i.e., $\omega = 1$, therefore a dissipative fluid with EoS $p = \rho$ is not allowed in this description. 

\section{A new solution of the Israel-Stewart equation} 
\label{sec:new}
Using the Eqs. (\ref{eq:eq18})-(\ref{eq:eq1}), we obtain the following differential equation also known as IS transport equation, 
\begin{widetext}
\begin{equation}
\ddot{H}+\left[3H(1+\omega)+\frac{\Delta}{2} \right]\dot{H}+\frac{9}{2}\epsilon \left(1-\omega^{2} \right)\left\lbrace \frac{(1+\omega)}{3^{s}\xi_{0}}H^{1-2s}-1 \right\rbrace H^{3} + \frac{\epsilon (1-\omega^{2})}{3^{s-1}\xi_{0}}\dot{H}H^{2(1-s)}+\frac{3}{4}(1+\omega)\Delta H^{2}=0.
\label{eq:is}
\end{equation}
\end{widetext}
As discussed in Ref. \cite{CruzLepe}, this equation admits a phantom solution for a late time FLRW flat universe filled with only one fluid described by a barotropic EoS and bulk viscosity. This was done for $s=1/2$, $\epsilon =1$, and the solution has the form
\begin{equation}
H\left( t\right) =A\left(t_{s}-t\right)^{-1}, 
\label{eq:Ansatz}
\end{equation}
which has a Big Rip singularity at a finite value of cosmic time $t_{s}$ in the future. More details about this phantom solution can be found in \cite{Cataldo2005, CruzLepe} within the framework of the IS theory and \cite{CCL}, where a non-linear extension of IS theory was considered. From the conservation equation (\ref{eq:eq18}), it can be observed that viscous effects are introduced as follows,
\begin{equation}
p_{eff} = p + \Pi,
\end{equation}
for a barotropic fluid one gets
\begin{equation}
\omega_{eff} = \omega + \frac{\Pi}{3H^{2}},
\label{eq:omegaeff}
\end{equation} 
where the Friedmann equation (\ref{eq:eq0}) was used. From the acceleration equation of motion, the viscous pressure (or dissipation term) takes the form
\begin{equation}
\Pi = -2\dot{H}-3(1+\omega)H^{2},
\label{eq:deviation}
\end{equation}
with this expression we quantify the deviations from equilibrium and it is expected to obtain the cosmic accelerated expansion from this negative effective pressure \cite{viscous}. According to Eq. (\ref{eq:eq2}), we have defined 
\begin{equation}
\Delta :=\frac{3H}{\delta \left( \omega \right) }\left( \delta \left(\omega \right) -\frac{\dot{H}}{H^{2}}\right), 
\label{eq:eq4}
\end{equation}
for simplicity in the notation we define $\delta (\omega) := (3/4)\left[(1+\omega)/(1/2+\omega)\right]$.\\

In this work we will also take $s=1/2$, since, as we mentioned above, the simple Ansatz given by Eq. (\ref{eq:Ansatz}) represents a phantom solution of Eq. (\ref{eq:is}), which for other values of $s$ it becomes in a very difficult equation to solve analytically. Moreover, for $s$ different from $1/2$ one can find numerically that no other phantom solution exist. From this point of view, the case $s=1/2$ yields the possibility to have behaviors not present with other values of $s$.\\   
Strictly speaking, the coefficients of viscosity must be evaluated from the kinetic theory but in general their EoS and thermodynamics coefficients are very complicated. In order to find analytical solutions, simplified dependence between the coefficient of viscosity and the fluid density are assumed, as we do in this work. Despite of this simplicity, we can gain a deep insight of the physical properties of cosmic fluids with dissipation. Therefore we will try to address the current dilemma of cosmic evolution with the following Ansatz for Hubble parameter in the IS equation
\begin{equation}
H(t>t_{s}) = \frac{\left| A\right|}{t_{s}}\left(\frac{t}{t_{s}}-1 \right)^{-1}.
\label{eq:quinte}
\end{equation}
As we will see below this brand-new form for the Hubble parameter will unfold some attractive characteristics from a cosmological point of view of the IS theory. This new solution posses also an initial singularity at the time $t \approx t_{s}$, which is a common property of the Friedmann models with barotropic EoS. From now on, we will focus on the cosmological consequences of this new solution and its thermodynamical properties.\\

\begin{figure}[htbp!]
\centering
\includegraphics[width=6cm,height=4.5cm]{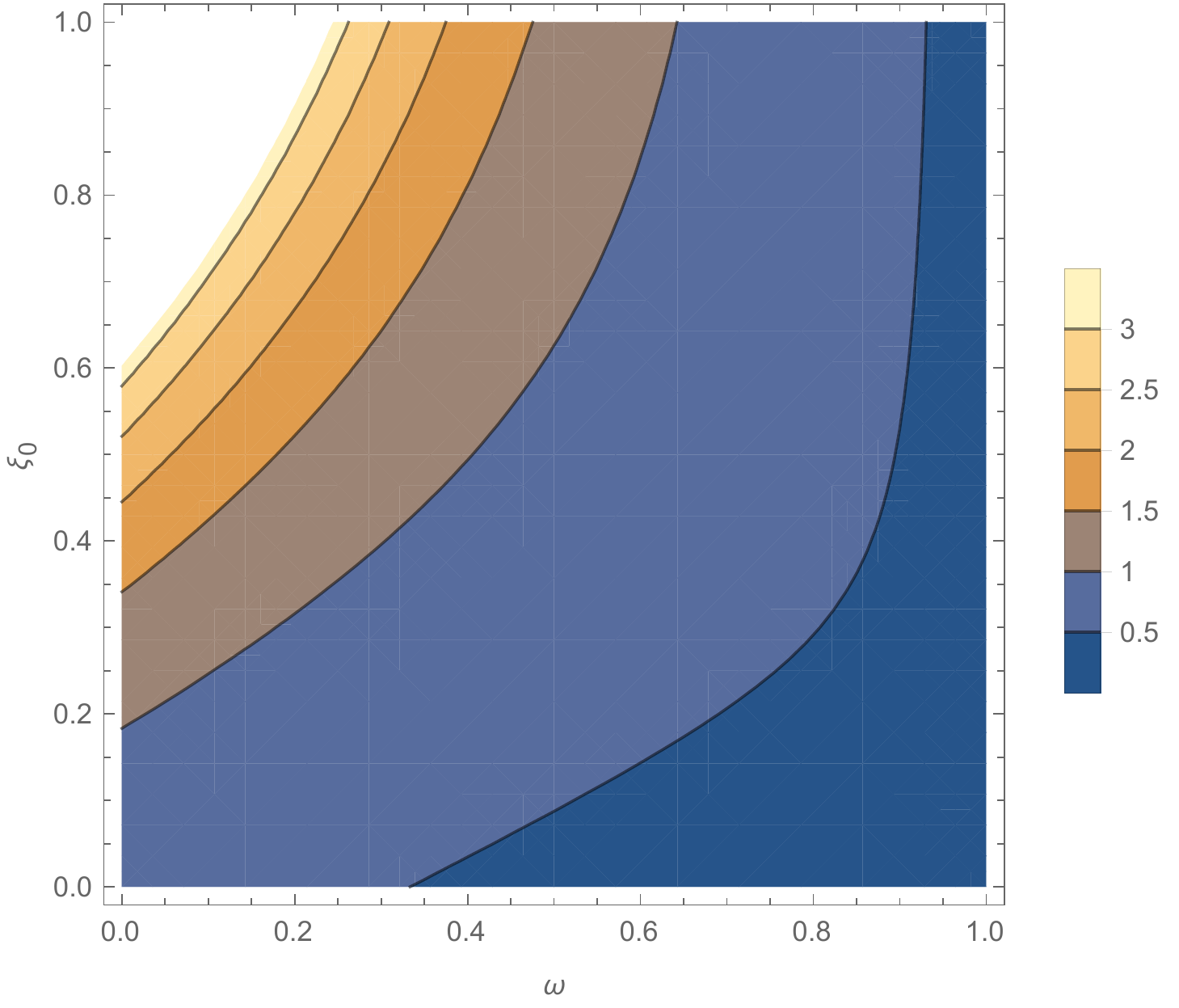}
\includegraphics[width=6cm,height=4.5cm]{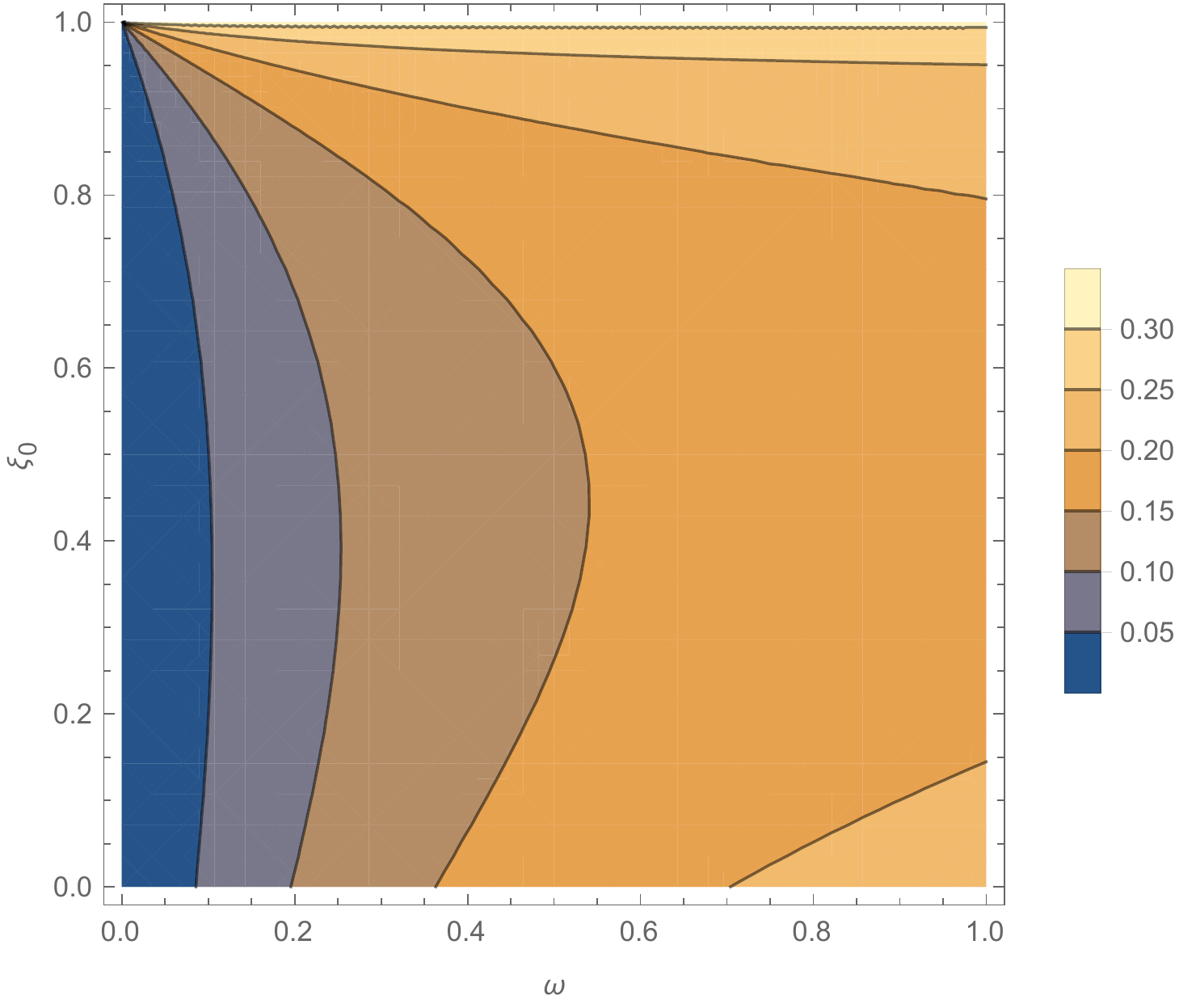}
\includegraphics[width=6cm,height=4.5cm]{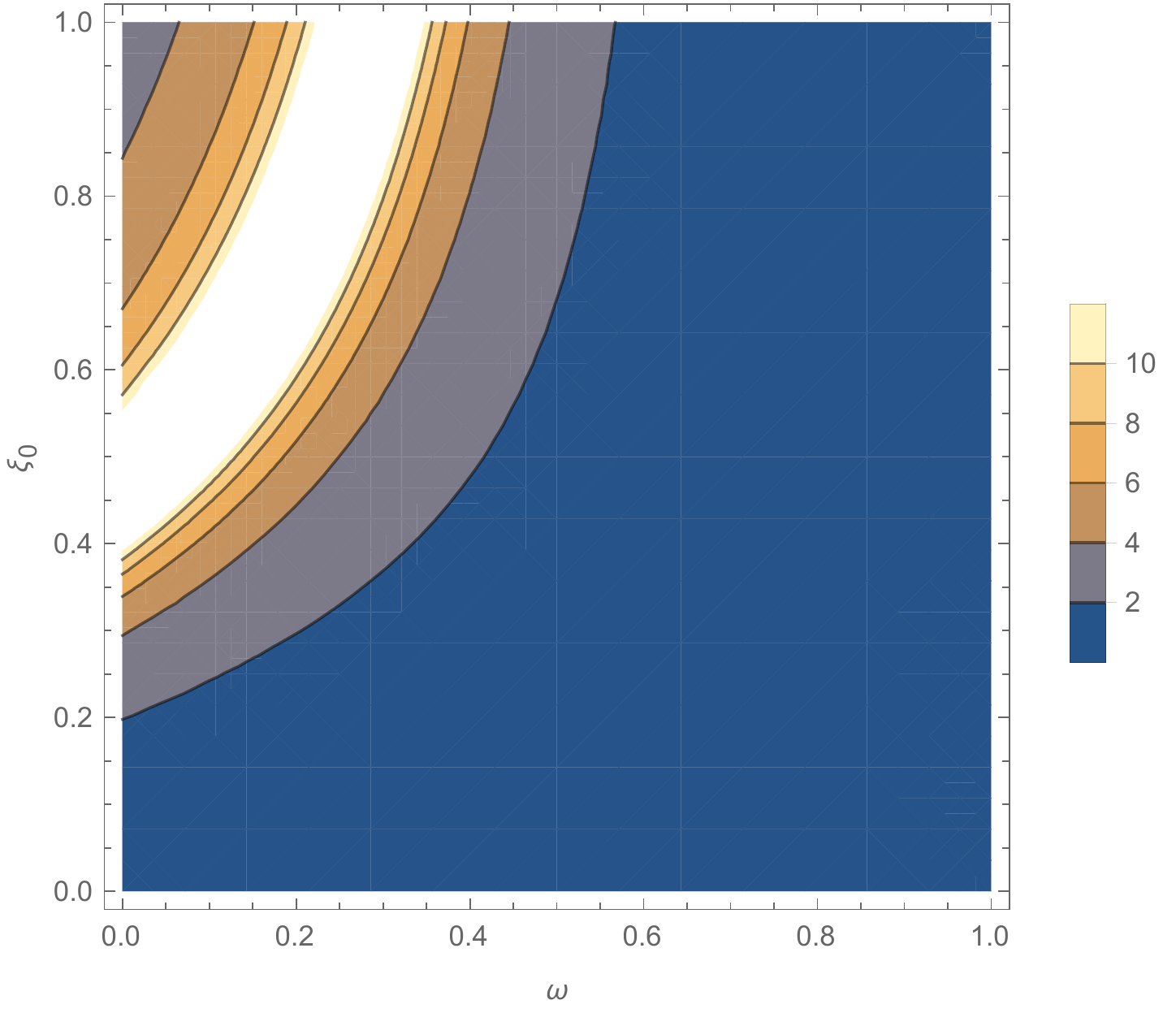}
\includegraphics[width=6cm,height=4.5cm]{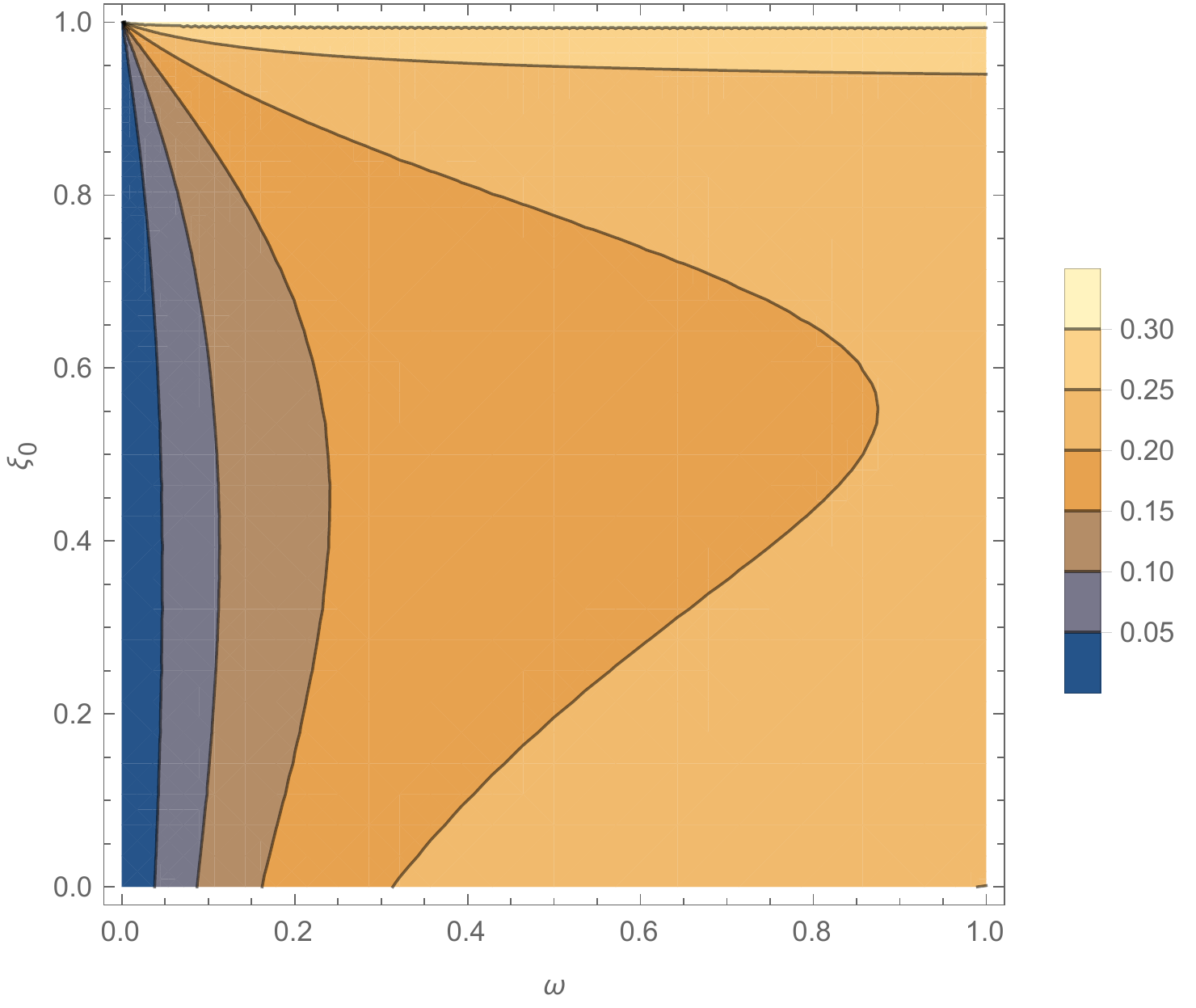}
\includegraphics[width=6cm,height=4.5cm]{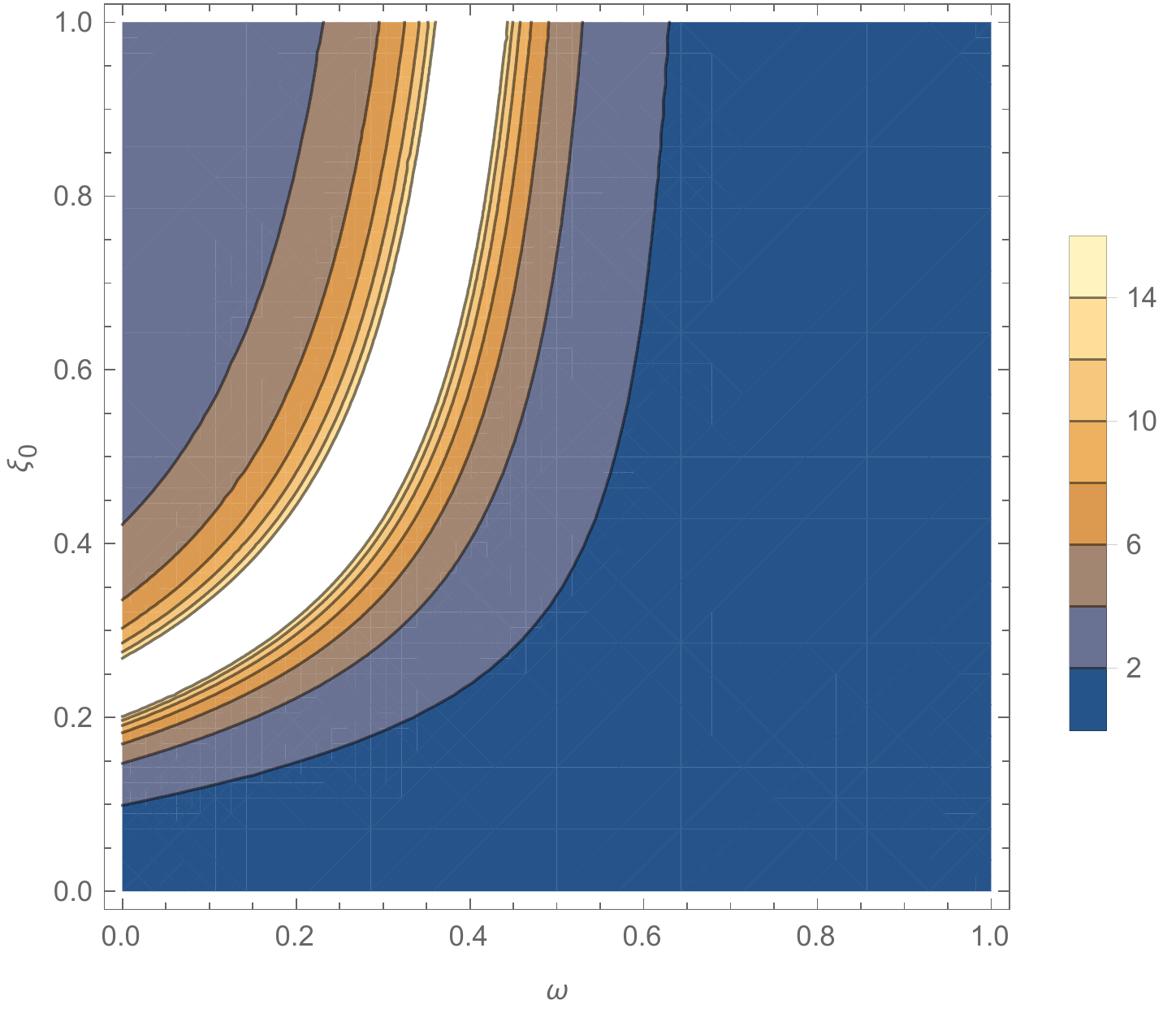}
\includegraphics[width=6cm,height=4.5cm]{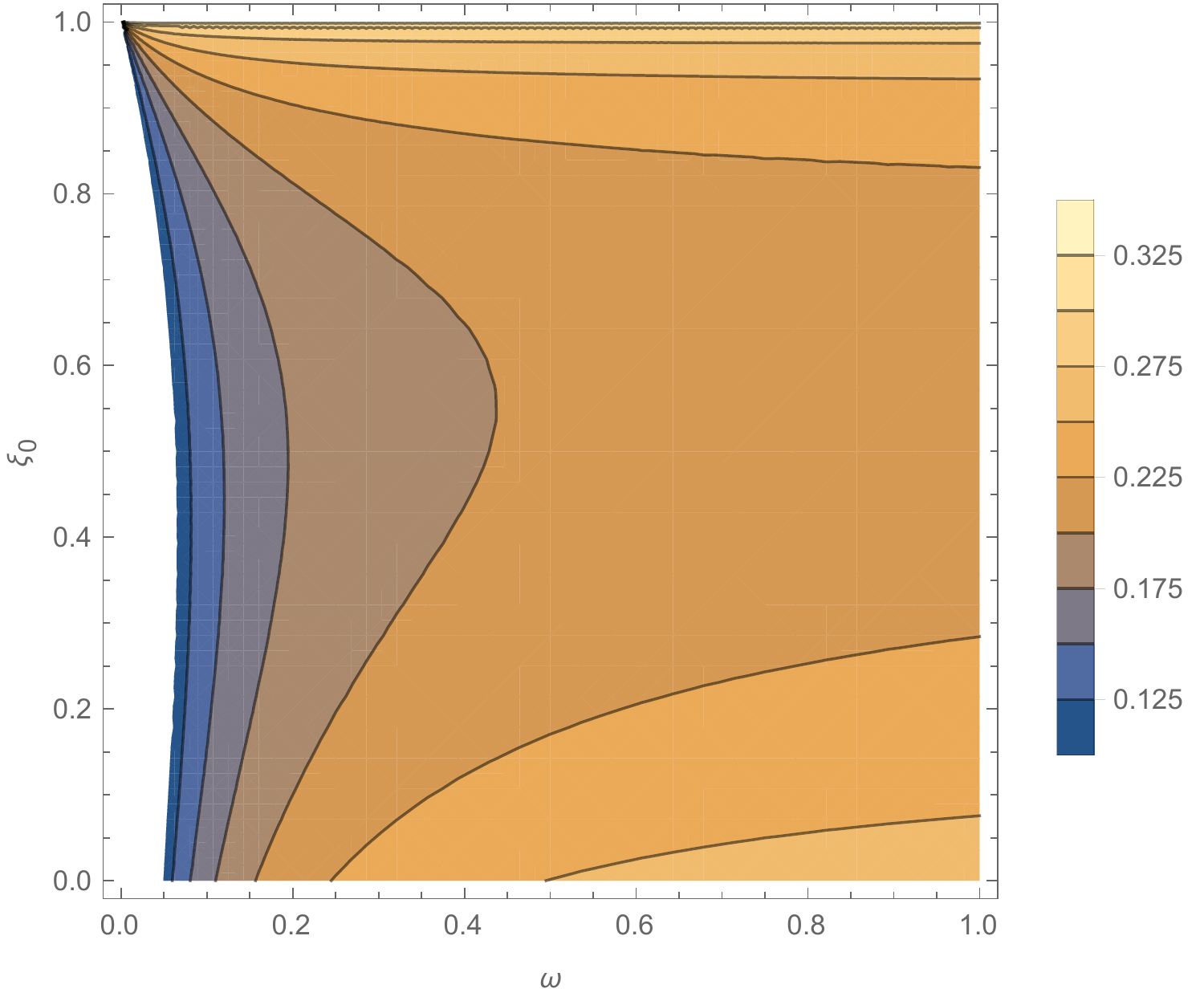}   
\caption{Behavior of branches $\left| A\right|_{\pm}$. The upper panels were obtained with $\epsilon = 0.9$, for the middle panels we used $\epsilon = 0.4$ and in the bottom panels we used $\epsilon = 0.2$. The right column represents the $A_{+}$ branch while the left column to $A_{-}$.} 
\label{fig:absolution}
\end{figure}

Once this new expression for $H(t)$ is introduced into the transport equation of the IS theory (\ref{eq:is}) for the specific case $s = 1/2$, we must verify that in fact it is solution. We obtain a quadratic equation for $\left|A\right|$ with two positive and real solutions. Since $\left|A\right|$ will always be positive, the Hubble parameter will be that of an expanding universe, after a straightforward calculation we are left with a quadratic equation for $\left|A\right|$ with constant coefficients 
\begin{widetext}
\begin{align}
& 3\left[\epsilon (1+\omega)^{2}(1-\omega)-\sqrt{3}\epsilon \xi_{0}(1-\omega^{2})+\frac{\sqrt{3}\xi_{0}}{2}(1+\omega) \right]\left| A\right|^{2}+\left[\sqrt{3}\xi_{0}\left(\frac{3}{2\delta(\omega)}-2 \right)(1+\omega)-\sqrt{3}\xi_{0}-2\epsilon(1-\omega^{2})\right]\left| A\right| \nonumber \\ 
& + \frac{4\xi_{0}}{\sqrt{3}}\left(1-\frac{3}{4\delta(\omega)} \right)=0,
\label{eq:absolute}
\end{align}
\end{widetext}
the roots for this equation are given by
\begin{widetext}
\begin{equation}
\left|A\right|_{\pm} = \frac{2 \left(\epsilon \omega (\omega + \omega^{2}-1)-\epsilon -\sqrt{3}\xi_{0}(1+\omega) \pm \sqrt{[6\xi^{2}_{0}(1+\omega)-\omega ^4 \epsilon ^2](1-\omega^{2})+2
   \omega ^5 \epsilon ^2-4 \omega ^3 \epsilon ^2-\omega ^2 \epsilon ^2+2 \omega  \epsilon ^2+\epsilon ^2} \right)}{3 \left(-2 \sqrt{3} \xi_{0}  \omega ^3-3 \sqrt{3} \xi_{0}  \omega ^2+\sqrt{3} \xi_{0} +2 \omega ^4 \epsilon +4 \omega ^3 \epsilon -4 \omega  \epsilon -2 \epsilon \right)}, 
\end{equation}
\end{widetext}
then, $\left|A\right|_{\pm} = \left|A\right|_{\pm}(\omega, \xi_{0}, \epsilon)$. Hereafter we will consider the following condition $0 \leq \omega < 1$ to avoid a singular behavior in the relaxation time, $\tau$, and we will restrict ourselves on the interval $[0,1]$ for the parameter $\xi_{0}$. In order to preserve the causality condition the parameter $\epsilon$ must be restricted to the interval $0< \epsilon \leq 1$. Using the intervals mentioned before for each parameter we can find two real solutions for $\left| A \right|_{\pm}$. In Fig. (\ref{fig:absolution}) we can visualize the branches $\left| A\right|_{\pm}$ in the space of parameters $(\omega, \xi_{0})$ and some fixed values of $\epsilon$. By varying the parameter $\epsilon$ we have different cases for each branch of the solution, however, only one branch of the solution $\left|A\right|$ can take values greater (or less) than 1 while the second is always less than 1, as we will see below this second branch of the solution must be discarded. The fact that the solution can be greater or less than 1 is an important feature, since depending on what is the case, the cosmic evolution will have a different behavior, we will comment this later. 
\onecolumngrid

\twocolumngrid


\subsection{Initial singularity}
From Eq. (\ref{eq:quinte}) we can easily compute the curvature scalar $R=-6(\dot{H}+2H^{2})$, we obtain
\begin{equation}
R = -12\left(\frac{\left| A\right|}{t_{s}}\right)^{2}\left(1-\frac{1}{2\left| A\right|}\right)\left(\frac{t}{t_{s}}-1\right)^{-2},
\label{eq:curvature}
\end{equation}
where we have chosen the value $k = 0$ (flat geometry), for the scale factor we have
\begin{equation}
a(t) = a_{0}\left(\frac{t-t_{s}}{t_{0}-t_{s}} \right)^{\left| A\right|},
\label{eq:scale}
\end{equation} 
when $t \approx t_{s}$ we have simultaneously that $R \rightarrow \infty$ and $a\rightarrow 0$, this behavior has the characteristics of a Big Bang singularity, in this case $t_{s}$ represents a time in the past in which the singularity took place. From the expression for the scale factor, the redshift can be expressed as
\begin{equation}
1+z =  \left(\frac{t_{0}-t_{s}}{t-t_{s}} \right)^{\left|A\right|},
\label{eq:red} 
\end{equation}
note that for $t \approx t_{s}$ we have $z\rightarrow \infty$ for all $\left|A\right|$, which corresponds to the origin of the universe. On the other hand, as $t$ increases we will have $z \rightarrow -1$. By introducing the deceleration parameter we have
\begin{equation}
1+q = -\frac{\dot{H}}{H^{2}} = \frac{1}{\left| A\right|}.
\label{eq:decel}
\end{equation}
According to the standard cosmological model (SCM), for a spatially flat universe modelled by a single fluid and barotropic EoS, the deceleration parameter takes the form
\begin{equation}
q = \frac{1}{2}(1+3\omega), 
\end{equation}
since $a(t) \propto t^{2/3(1+\omega)}$ we must have the condition $-1 \leq \omega \leq 1$ in order to avoid superluminal velocities. For these values on the parameter $\omega$ one easily gets that $-1 \leq q \leq 2$. Then to be within the limits established by the SCM for the deceleration parameter, in our approach we must demand $\left|A\right| \geq 1/3$. From Fig. (\ref{fig:absolution}) we can see clearly that previous constraint for $\left|A\right|$ is not satisfied by the branch $\left|A\right|_{+}$, therefore is not considered in our analysis, as we will see later this argument is not unique to discard this branch of the solution, some thermodynamic criteria are not favored by this branch either. It is worthy to mention that branch $\left|A\right|_{-}$ can take values less than 1 but always greater than 1/3.\\

In the SCM it is possible to write $q = q(z)$, for the early universe $q(z\rightarrow \infty) = 1/2$, therefore, from Eq. (\ref{eq:decel}) we have that in this case $\left|A\right| = 2/3$. On the other hand, for a dominated cosmological constant expanding universe $q(z\rightarrow -1) = -1$ is obtained in our approach only when $\left|A\right| \rightarrow \infty$, finally for radiation we have $q = 1$ which implies $\left|A\right| = 1/2$. In Ref. \cite{staro} was found that for an evolving dark energy model the best fit for the deceleration parameter was $q_{0} = -0.63 \pm 0.12$ at present, this was done using the supernova + WMAP CMB data, in order to be in agreement with this dark energy model, in our approach we must demand $\left|A \right| \geq 2.04$ for the upper bound and $\left|A \right| \geq 4$ for the lower.\\ Therefore, by imposing the limit $\left|A\right| \geq 1/3$ we can have specific values for the deceleration parameter that can emulate some stages of the SCM and at same time we can be observe that for higher values of the solution $\left|A\right|$ we can reach for the deceleration parameter similar values as those obtained in the aforementioned dark energy model.

\subsection{Accelerated expansion: $\left|A\right| > 1$}
In terms of the deceleration parameter the curvature scalar can be written as $R = -6H^{2}(1-q) = 3H^{2}(1+3\omega_{eff})$, for an accelerated epoch $q < 0$, the scalar $R$ remains negative, by considering the parameter $\omega_{eff}$ the negativity of $R$ can be maintained with a quintessence or phantom fluids, according to \cite{wmap} we can have $\omega(0)=-1.019_{-0.080}^{+0.075} = -0.944$ which corresponds to quintessence zone and $\omega(0) = -1.099 \approx -1.1$ for the phantom zone, recent evidence coming from the DES collaboration situates the parameter $\omega(0)$ between -1.3 and -0.56 \cite{des}. Otherwise, for a decelerated universe $q > 0$, the curvature scalar could turn positive. This change in the sign of the curvature scalar can be also obtained if we consider $\left|A\right| < 1/2$ or $\left|A\right| > 1/2$, respectively, see Eq. (\ref{eq:curvature}).\\
  
In order to have an accelerated expansion we must demand $\ddot{a}(t) > 0$, from Eqs. (\ref{eq:scale}) and (\ref{eq:decel}) we can write
\begin{eqnarray}
\ddot{a}(t) &=& a_{0}\left| A\right| \left(\left| A\right| - 1\right)\frac{\left(t-t_{s}\right)^{\left| A\right|-2}}{(t_{0}-t_{s})^{\left| A\right|}}, \label{eq:accel}\\ 
&=& -a_{0}\left| A\right|^{2}q\frac{\left(t-t_{s}\right)^{\left| A\right|-2}}{(t_{0}-t_{s})^{\left| A \right|}},
\end{eqnarray}  
then, $\left| A\right| > 1$ or $q < 0$ represents an accelerated expansion. Using the Ansatz given in Eq. (\ref{eq:quinte}) we can write the viscous pressure (\ref{eq:deviation}) as follows
\begin{equation}
\Pi(t) = -\left(\frac{\left|A\right|}{t_{s}}\right)^{2}\left[3(1+\omega)-\frac{2}{\left|A\right|}\right]\left(\frac{t}{t_{s}}-1 \right)^{-2}.
\label{eq:devifrw}
\end{equation}
Note that near the Big Bang, $t \approx t_{s}$, the viscosity effects (or temperature, see Eq. (\ref{eq:temp})) are very high. Such effects are physically viable since all the fields are confined in a small region of the spacetime. As discussed in Ref. \cite{SdC} within the context of warm inflation, the viscosity effects are of great relevance for entropy producing systems and high rate multi-particle interaction. On the other hand, far from Big Bang, $t \gg t_{s}$, the viscosity (and temperature, see Eq. (\ref{eq:temp})) dilutes. This behavior has certain similarities with the characteristics of the early universe for times after the Big Bang. Once the expansion and cooling were given, the universe consisted of a ``soup'' composed by quarks, gluons, electrons and neutrinos, called quark-gluon plasma (QGP). This QGP has the characteristic of presenting viscosity (in the same sense as for the fluid used here: resistance to expansion or contraction) and some advances in the understanding of its properties were obtained by recreating it with some particle collisions experiments, with these experiments was determined that its viscosity could be low \cite{qgp}, however, a more precise measuring of this viscosity is still a subject of research posed by experimental collaborations in the particle physics field.\\ 

The negativity of the viscous pressure is preserved for the following cases $\left|A\right| > 1$ or $2/3(1+\omega) < \left|A\right| < 1$, as can be observed from Eq. (\ref{eq:devifrw}). Besides, from Eqs. (\ref{eq:omegaeff}), (\ref{eq:deviation}), (\ref{eq:quinte}) and (\ref{eq:decel}), we have the following expression 
\begin{equation}
\omega_{eff} = -1-\frac{2}{3}\frac{\dot{H}}{H^{2}} = -1+\frac{2}{3}(1+q) = -1+\frac{2}{3}\frac{1}{\left|A\right|}.
\label{eq:effective}
\end{equation}
Note that with the condition $\left|A\right| > 1 \ (\mbox{or} - 1 \leq q < 0 )$, the accelerated expansion is due to a quintessence fluid and began in the past with a singularity. From this last equation  for $\omega_{eff}$ we observe that a phantom behavior during the evolution is not allowed in this description.

\subsection{Milne universe}
From Eq. (\ref{eq:accel}) we can see that $\ddot{a} = 0$ when $\left|A\right| = 1$, i.e., not accelerated cosmic expansion, alternatively from Eq. (\ref{eq:scale}) we can have a direct evaluation of the first derivative with respect to time of the scale factor with $\left|A\right| = 1$ yielding $\dot{a} = a_{0}(t_{0}-t_{s})^{-1}$, which is a constant value. For this value of $\left|A\right|$ the deceleration parameter expressed in Eq. (\ref{eq:decel}), takes the value $q = 0$.\\ 
It is worthy to mention that if we consider $1< \left|A\right| < 2$ we can have the following behavior for the acceleration, $\ddot{a}(t\rightarrow \infty) \rightarrow 0$, but the deceleration parameter $q$ may take several values ​​and always different from zero for any time, this situation does not represent a Milne evolution.\\ 
For the Milne evolution the curvature scalar is always negative, see Eq. (\ref{eq:curvature}), and the effective parameter $\omega_{eff}$ has a constant value given by $-1/3$. The negativity of the viscous pressure, $\Pi(t)$, is always guaranteed. 

\subsection{Decelerated expansion: $\left|A\right| < 1$}
\begin{figure}[H]
\centering
\includegraphics[scale=0.58]{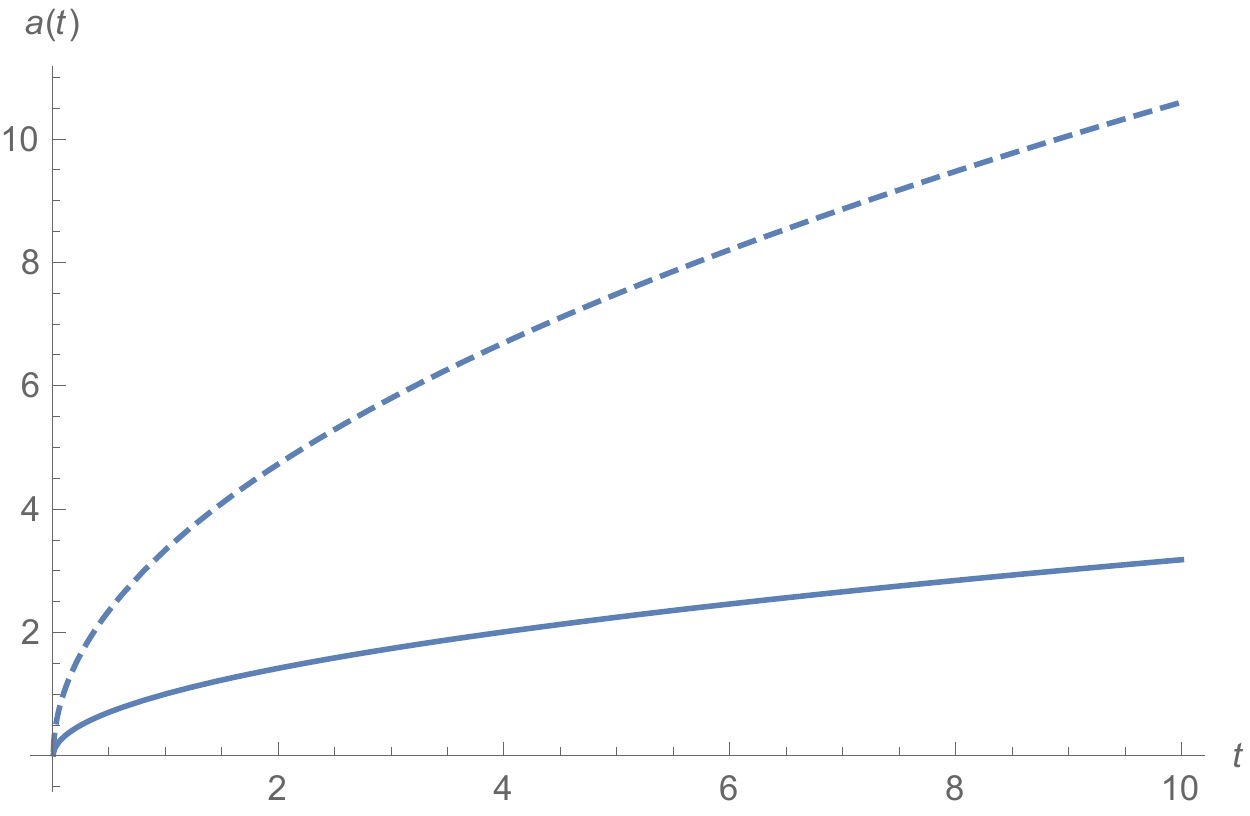}
\caption{Behavior of scale factor for $\left|A\right| = 1/2$ with the conditions $t_{s} \gg t_{0}$ (dashed line) and $t_{s} \ll t_{0}$ (solid line).} 
\label{fig:scaleunmedio}
\end{figure}

In this case we are out of the quintessence zone and the cosmic expansion is decelerated, this can be seen from Eq. (\ref{eq:accel}) where $\ddot{a} < 0$ when $\left|A\right| < 1$. By re-expressing the Eq. (\ref{eq:effective}) as follows
\begin{equation}
\omega_{eff} = -1-\frac{2}{3}\frac{\dot{H}}{H^{2}} = -\frac{1}{\left|A\right|}\left(\left|A\right|-\frac{2}{3}\right),
\label{eq:effective2}
\end{equation}
we obtain two possibilities for the effective parameter $\omega_{eff}$, $\omega_{eff} < 0$ if $2/3 < \left|A\right| < 1$ and $\omega_{eff} > 0$ when $ 1/3 \leq \left|A\right| < 2/3$. From these limits the quintessence (or phantom) zone is forbidden.\\ 
It is worth noting that in the case $\left|A\right| = 1/2$ we have a null value for the scalar curvature and $\omega_{eff} = 1/3$. In Fig. (\ref{fig:absolution}) we can observe that this limit can be obtained by considering $\epsilon = 0.9$ (upper panel on the left), for instance. According to Fig. (\ref{fig:scaleunmedio}) we can observe the behavior of the scale factor expressed in Eq. (\ref{eq:scale}) and $\left|A\right| = 1/2$, if we consider $t_{s} \ll t_{0}$ the growth of the scale factor is slower (solid line). This result for the scale factor presents a similar behavior as the one obtained for a universe dominated by radiation (relativistic gas of photons and neutrinos) in standard cosmology where $\rho \propto a^{-4}$. 

\subsection{Age of the universe}  
It is well-established that the universe had a decelerated expansion phase once the inflation process stopped, however after some time the universe presented a transition towards a phase of accelerated expansion, we will give an estimation of the age of the universe at the moment of this transition. Using the Eqs. (\ref{eq:quinte}) and (\ref{eq:scale}), we can write the Hubble parameter as follows
\begin{equation}
H(a) = \frac{\left|A\right|}{(t_{0}-t_{s})}\left(\frac{a}{a_{0}}\right)^{-1/\left|A\right|},
\end{equation} 
note that $\left|A\right|$ must be restricted to the values of a decelerated universe $1/3 \leq \left|A\right| \leq 1$. We can perform the integration \cite{transition}
\begin{equation}
t(a_{t}) = \int^{a/a_{t}}_{0}\frac{da}{aH(a)} = (t_{0}-t_{s})\left(\frac{a_{t}}{a_{0}}\right),
\end{equation}
where we can assume $t_{s} \approx 0$, therefore we can have an estimation of the age of the universe at deceleration-acceleration transition
\begin{equation}
\mbox{Age}_{t} := t_{0}\left(\frac{a_{t}}{a_{0}}\right),
\label{eq:age}
\end{equation} 
we are denoting by the subindex $t$ the transition moment. In terms of the redshift (\ref{eq:red}) we can write for the Hubble parameter (\ref{eq:quinte})
\begin{equation}
H(z) = \frac{\left|A\right|}{t_{0}-t_{s}}(1+z)^{1/\left|A\right|},
\end{equation} 
and using $1+z = a_{0}/a$ we have $da/a = - dz/(1+z)$, then the integral (\ref{eq:age}) results
\begin{equation}
t(z_{t}) = (t_{0}-t_{s})(1+z_{t})^{-1/\left|A\right|}, 
\end{equation} 
therefore
\begin{equation}
\mbox{Age}_{t} := t_{0}(1+z_{t})^{-1/\left|A\right|},
\end{equation}
where $t_{s} \approx 0$. The value $z_{t}$ has been estimated in several works providing $z_{t} = 0.82 \pm 0.08$ using BAO data \cite{bao}. From the compilation of 28 independet measurements of the Hubble parameter $z_{t}$ was estimated as $z_{t} = 0.74 \pm 0.05$ \cite{28} and for an updating of the aforementioned compilation $z_{t}$ was obtained as $z_{t} = 0.72 \pm 0.05 \ (0.84 \pm 0.03)$ for $H_{0}= 68 \pm 2.8 \ (73.24 \pm 1.74) \ \mbox{(km/s)Mpc}^{-1}$ in the framework of dark energy models \cite{38}. The actual age of the universe, $t_{0}$, is estimated around 14 Gyr \cite{highz, age} and its age at transition moment around 9 Gyr \cite{transition}. Since the solution $\left|A\right|$ is restricted on the interval $1/3 \leq \left|A\right| \leq 1$ then $1 \leq \left|A\right|^{-1} \leq 3$, we will have
\begin{equation}
\mbox{Age}_{t,min} = t_{0}(1+z_{t})^{-3}, \ \ \mbox{Age}_{t,max} = t_{0}(1+z_{t})^{-1},
\end{equation}
being $\mbox{Age}_{t,max}$ more accurate. The best estimation for the age of the universe is obtained with $z_{t} = 0.67$, which is the lower bound coming from $z_{t} = 0.72 \pm 0.05$ given in Ref. \cite{38} yielding $\mbox{Age}_{t,max} \approx 8.4$ Gyr.\\  

\section{Thermodynamical properties of the new Israel-Stewart solution}
\label{sec:thermo}  
By considering the integrability Gibbs condition, we can establish for the temperature
\begin{equation}
T(t) = T_{0}\rho^{\omega/(1+\omega)} = T_{0} \left[3\left(\frac{\left| A\right|}{t_{s}}\right)^{2}\left(\frac{t}{t_{s}}-1 \right)^{-2}\right]^{\omega/(1+\omega)},
\label{eq:temp}
\end{equation}
where the Friedmann equation was used, as expected, near the Big Bang given by the limit $t\rightarrow t_{s}$, the temperature has a singular behavior. If we consider that number of particles is conserved, we have the following continuity equation
\begin{equation}
\dot{n}+3Hn = 0,
\end{equation}  
where $n$ is the number density of particles, by direct integration using the Ansatz (\ref{eq:quinte}), one gets
\begin{equation}
n(t) = n(t_{0})\left(\frac{t_{0}}{t_{s}}-1 \right)^{3\left|A\right|}\left(\frac{t}{t_{s}}-1 \right)^{-3\left|A\right|}
\end{equation}
from this expression we get that for $n(t \rightarrow t_{s}) \rightarrow \infty$, as expected since in the Big Bang singularity the volume $V \rightarrow 0$. For the case $t \gg t_{s}$ we can observe that this density decreases. For a fluid the near equilibrium condition must be
\begin{equation}
\left|\Pi\right| \ll p,
\label{eq:near}
\end{equation}
using the Eqs. (\ref{eq:deviation}) and (\ref{eq:decel}) we can compute the quotient
\begin{equation}
\left|\frac{\Pi}{p}\right| = \frac{1}{3\omega}\left|3(1+\omega)-\frac{2}{\left|A\right|} \right|,
\end{equation}
by considering $\left|\Pi / p \right| \ll 1$ we must have
\begin{equation}
\left|A\right| \ll \frac{2}{3}.
\label{eq:far}
\end{equation} 
As observed, this condition it is contradictory to the one coming from the SCM for the deceleration parameter which implies $\left|A\right| \geq 1/3$, then, the analysis presented here is not close to the equilibrium condition. Under the near equilibrium condition given by Eq. (\ref{eq:near}), the IS transport equation (\ref{eq:is}) can be written as
\begin{align}
& \ddot{H}+\underbrace{\left[3(1+\omega)+\frac{\sqrt{3}\epsilon(1-\omega^{2})}{\xi_{0}}\right]}_{\beta}H\dot{H} \nonumber \\ 
& + \underbrace{\left[\frac{3\sqrt{3}\epsilon (1-\omega^{2})(1+\omega)}{2\xi_{0}}-\frac{9}{2}\epsilon (1-\omega^{2})\right]}_{\alpha}H^{3} = 0,
\end{align}
which corresponds to the truncated IS theory \cite{maartens}. Inserting the Ansatz (\ref{eq:quinte}) in previous equation we can find a cubic equation with constant coefficients for $\left|A\right|$, which can be written as $\alpha \left|A\right|^{3}-\beta \left|A\right|^{2}+2\left|A\right| = 0$, whose solutions are given by $\left|A\right|_{1} = 0$ and $\left|A\right|_{2,3} = [\beta \pm \sqrt{\beta^{2}-8\alpha}]/(2\alpha)$. By considering the space of parameters $(\omega, \xi_{0})$ and some fixed values of the parameter $\epsilon$ we can find some regions where the condition (\ref{eq:far}) is valid in this new set of solutions for $\left|A\right|$ yielding thermodynamical consistency.\\
 
Using the previous results the entropy production expression
\begin{equation}
\frac{dS}{dt} = -\frac{3H\Pi}{nT},
\label{eq:dergen}
\end{equation}
can be calculated by direct substitution, yielding
\begin{equation}
\frac{dS}{dt} = \Gamma \left[3(1+\omega)-\frac{2}{\left|A\right|} \right]\left(\frac{t}{t_{s}}-1 \right)^{3(\left|A\right|-1)+2\omega/(1+\omega)},
\label{eq:first}
\end{equation}
where we have defined the cumbersome constant $\Gamma$ as 
\begin{eqnarray}
\Gamma &:=& 3^{1-\omega/(1+\omega)}(n_{0}T_{0})^{-1}\left|A\right|^{3-2\omega/(1+\omega)}t_{s}^{-(3-2\omega/(1+\omega))}\times \nonumber \\ 
& \times & \left(\frac{t_{0}}{t_{s}}-1\right)^{-3\left|A\right|}.
\end{eqnarray} 
From Eq. (\ref{eq:first}) we can compute the second derivative of the entropy, we have
\begin{eqnarray}
\frac{d^{2}S}{dt^{2}} &= \Gamma \left[3(1+\omega)-\frac{2}{\left|A\right|} \right]\left(3(\left|A\right|-1)+\frac{2\omega}{(1+\omega)} \right)\times \nonumber \\
& \times \left(\frac{t}{t_{s}}-1 \right)^{3(\left|A\right|-1)+2\omega/(1+\omega)-1}.
\label{eq:second} 
\end{eqnarray} 
Then, by integrating Eq. (\ref{eq:first}) we can obtain an expression for the entropy
\begin{equation}
S = \frac{t_{s}\Gamma}{\left|A\right|}(1+\omega)\left(\frac{t}{t_{s}}-1\right)^{3\left(\left|A\right| -1\right)+2\omega/(1+\omega)+1},
\label{eq:entropy}
\end{equation}
which is always positive. At thermodynamical level the evolution of natural processes demand two consistency conditions on the derivatives of the entropy, we must have $dS/dt > 0$ and the convexity condition given by $d^{2}S/dt^{2} < 0$. According to Eq. (\ref{eq:first}) the generation of entropy will be positive if
\begin{equation}
\frac{2}{3(1+\omega)} < \left|A\right|,
\label{eq:goodone} 
\end{equation}  
the l.h.s. of previous inequality is always less than 1. In order to satisfy the convexity condition for the entropy, from Eq. (\ref{eq:second}) we obtain
\begin{equation}
\left|A\right| < 1-\frac{2\omega}{3(1+\omega)},
\label{eq:sec}
\end{equation}
from these results we can see straightforwardly that in order to satisfy both conditions at same time we must have $\left|A\right| < 1$, i.e., only in a decelerated universe the conditions on entropy are satisfied. By reviewing the limit cases for the parameter $\omega$ if we consider $\omega = 0$ the condition $2/3 \leq \left|A\right| \leq 1$ satisfies simultaneously the conditions (\ref{eq:goodone}) and (\ref{eq:sec}) and for $\omega \approx 1$ we must have $1/3 < \left|A\right| < 2/3$.\\
 
On the other hand, for a Milne universe which is obtained when $\left|A\right| = 1$, we have from Eq. (\ref{eq:first}),
\begin{equation}
\frac{dS}{dt} > 0, \ \ \ \mbox{since} \ \ \ 3(1+\omega) > 2.
\end{equation}   
However, from Eq. (\ref{eq:second}) the convexity condition can not be fulfilled since the quantity $2\omega /(1+\omega)$ is always positive, therefore the conditions for thermodynamical consistency are not available in this case. As can be seen from the expression (\ref{eq:entropy}), the entropy production is always positive. Note that in this case the near equilibrium condition is not satisfied.\\ 

It is worthy to mention that a rapid (or super-rapid) expanding universe can not be described by using the near equilibrium condition within the IS theory or other versions of thermodynamical theories in general relativity, see Ref. \cite{maartens}, thus we must be aware of this equilibrium condition if our interest relies in the description of an expanding universe. As a matter of fact, when the near equilibrium condition is satisfied, the positivity of the first derivative of the entropy and its convexity must be demanded for every ordinary thermodynamical system \cite{callen}, as shown previously we are not close to this regime, therefore, we can not assume a strict accomplishment of the aforementioned conditions on entropy in our formulation.\\ 

\begin{figure}[htbp!]
\centering
\includegraphics[width=6cm,height=5cm]{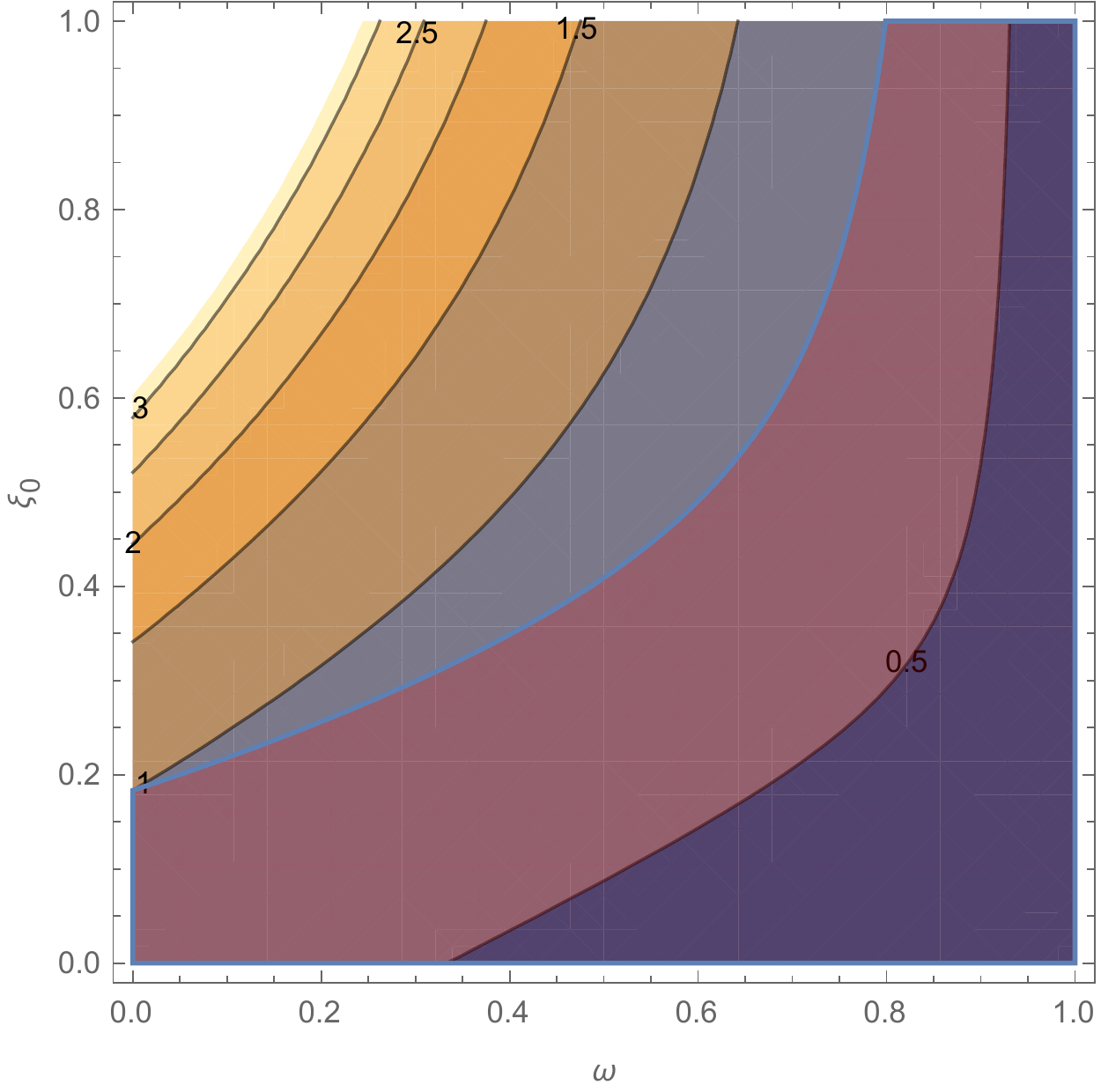}
\includegraphics[width=6cm,height=5cm]{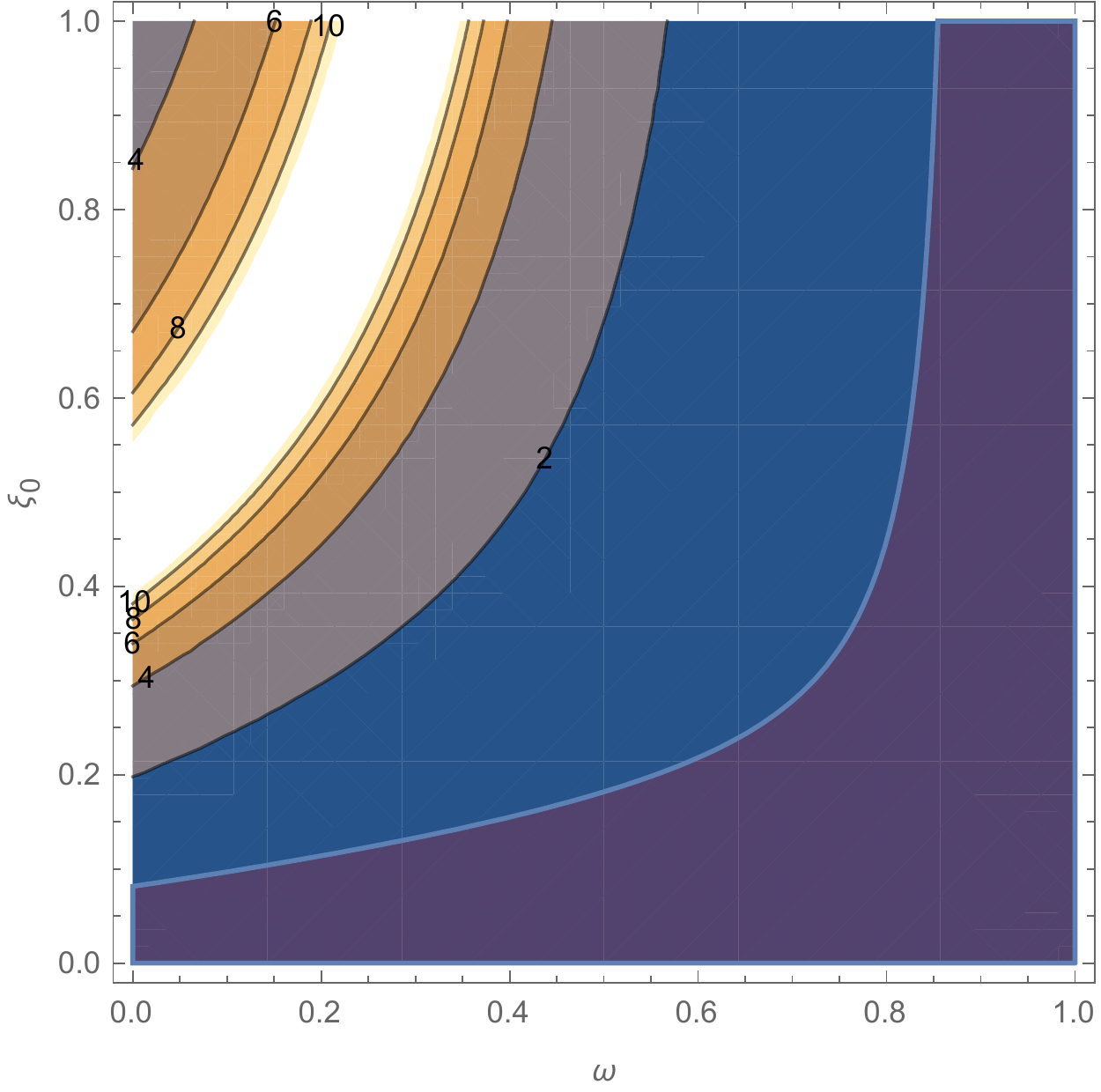}
\includegraphics[width=6cm,height=5cm]{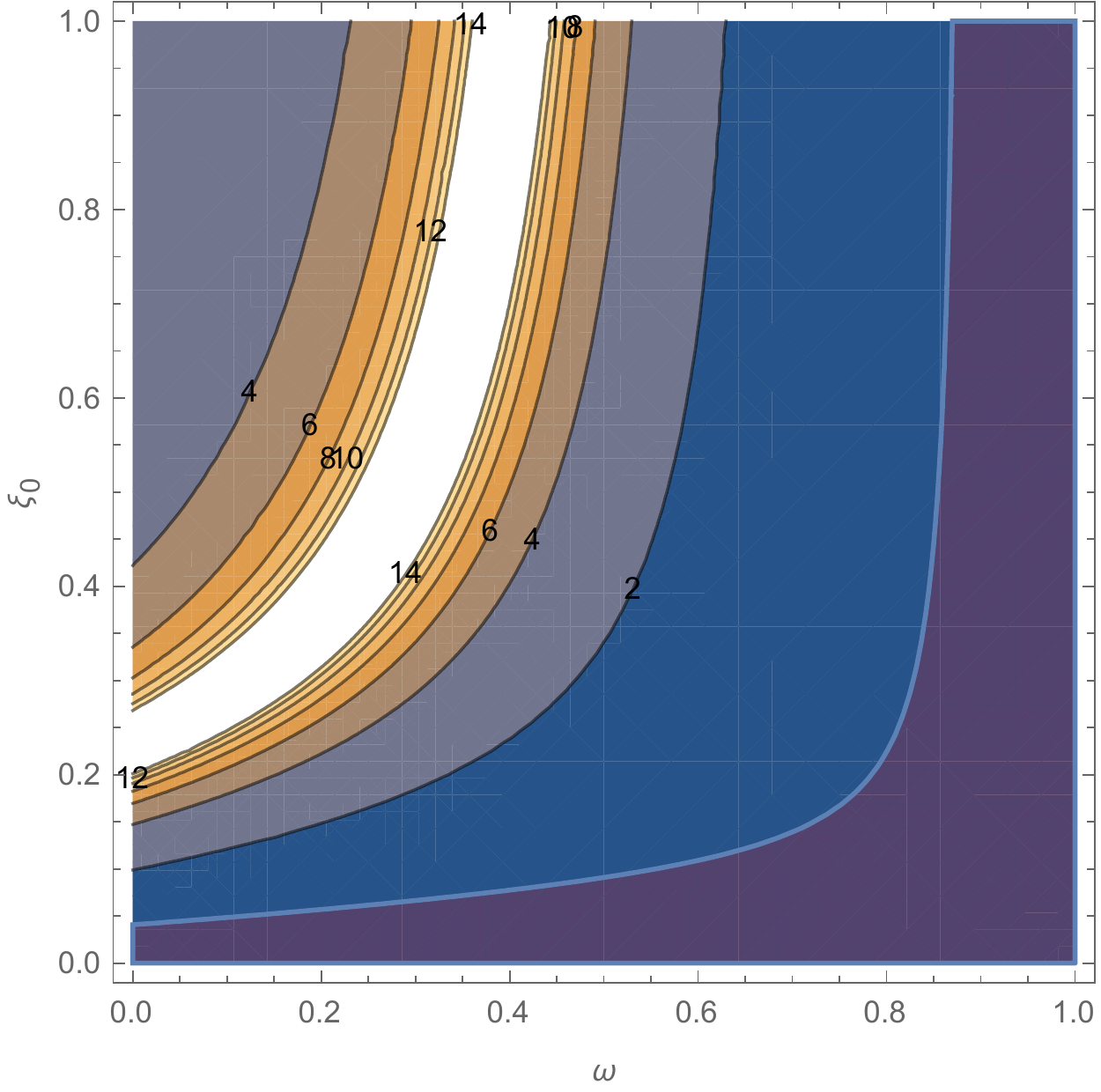}
\caption{In this plot we show the regions where $dS/dt > 0$ and the convexity condition $d^{2}S/dt^{2} < 0$ hold simultaneously in the branch $\left|A\right|_{-}$ shown in Fig. (\ref{fig:absolution}). The regions of thermodynamical consistency are shown in red.} 
\label{fig:entropy}
\end{figure}

As observed in Fig. (\ref{fig:entropy}), despite we are considering deviations from equilibrium by the introduction of the viscous pressure, for the branch $\left|A\right|_{-}$ there are regions in the space of parameters where these consistency conditions are guaranteed. For the second branch of the solution, $\left|A\right|_{+}$, there is not overlap of the regions generated by the conditions on the derivatives of the entropy given by Eqs. (\ref{eq:goodone}) and (\ref{eq:sec}), then is not thermodynamically consistent. Something similar was found for the non-linear extension of the IS theory in Ref. \cite{CCL}, by the inclusion of deviations and non-linear effects there are only some cases where the thermodynamical consistency is guaranteed.\\ 

In this description a late time acceleration of the universe could be driven by a quintessence-like fluid where $\left|A\right| > 1$ and in this case we always will have $d^{2}S/dt^{2} > 0$ together with $-1 < \omega_{eff} < -1/3$, a first consequence of this is an unbounded entropy, as can be seen from equation (\ref{eq:entropy}). This kind of behavior was found in \cite{termoPavon} where it is stated that convexity criterion for entropy is not always satisfied for expanding models with dominant fluids at late times, in other words, fluids driving the acceleration. In this approach the universe is modelled by a standard fluid with dissipative effects. Also in Ref. \cite{holo} was established that for a quintessence dominated era the generalized second law of thermodynamics ($dS/dt > 0$) can not be satisfied under the holographic scheme.\\ 

A second consequence on this fluid for not satisfying the convexity condition is that it will present certain instabilities that will be reflected as local inhomogeneities of temperature or matter \cite{callen}.\\

\subsection{Early universe} 
In what follows we discuss how our model with a viscous matter can mimic the behavior of stiff matter, radiation and dark matter in the framework of a fluid component in the usual cosmology. According to
the SCM the cosmic evolution of one fluid with a barotropic EoS, $p=\omega\rho$, is characterized by the following behavior for the energy density and the scale factor factor as a function of time
\begin{eqnarray}
\rho(a) &=& \rho_{0}\left(\frac{a}{a_{0}} \right)^{-3(1+\omega)},\\
a(t) &=& a_{0}\left[1+\frac{3(1+\omega)}{2}\sqrt{\frac{\rho_{0}}{3}}(t-t_{0}) \right]^{\frac{2}{3(1+\omega)}}.
\end{eqnarray}
In the case of stiff matter, $\omega \approx 1$, we have $ a(t) \sim (t-t_{0})^{1/3}$. In the case of radiation, $\omega= 1/3$, $ a(t) \sim(t -t_{0})^{1/2}$, and for dark matter, $\omega= 0$, $ a(t) \sim (t
-t_{0})^{2/3}$. By a simple comparison of these expressions with the scale factor given in Eq. (\ref{eq:scale}), we can infer that each case is obtained when $|A| \sim 1/3$, $|A| \sim 1/2$, and $|A| \sim 2/3$, respectively. In our description these values can be reached, for instance $|A| \sim 1/3$, if we take $\epsilon = 0.9$, $\omega = 0.9999$ and $\xi_{0} = 0.99$, for $|A| \sim 2/3$ we could take $\epsilon = 0.9$,  $\omega = 0.6$ and $\xi_{0} = 0.38$. These choices are not unique, both values for $|A|$ together with $|A| \sim 1/2$ are present in the solution shown in Fig. (\ref{fig:absolution}). If we consider the thermodynamical conditions given by Eqs. (\ref{eq:goodone}) and (\ref{eq:sec}), it is straightforward to see that the behavior of dust can be mimicked, for example, by a dissipative radiative fluid (see Fig.(\ref{fig:entropy})) with an $\xi_{0}$ in same range, which depends on the $\epsilon$ chosen. In this case, the negative pressure that comes from the dissipation inside the fluid leads to an effective zero pressure. To mimic the other cases we need to violate the thermodynamical conditions just mentioned.

\subsection{Near the stiff matter regime}
This case is obtained when $\omega \approx 1$, then the causality condition will be too close to its upper bound, $V^{2}= c_{s}^{2}+c_{b}^{2} \approx \omega$, where $c_{s}^{2}=\omega$ for a barotropic EoS and we assumed $c_{b}^{2}=\epsilon \left(1-\omega \right) \ll 1$. As commented before, in this regime the adjusting time for the falling temperature as the universe expands will take large values, this situation is not the desired one. The equation (\ref{eq:absolute}) can be approximated as 
\begin{equation}
\frac{3}{2}(1+\omega)\left|A\right|^{2}-2\left|A\right|+\frac{2}{3(1+\omega)} = 0,
\end{equation}  
which has the solution 
\begin{equation}
\left|A\right| = \frac{2}{3(1+\omega)} \simeq \frac{1}{3}.
\label{eq:stiff}
\end{equation}
This represents a decelerated universe with $\omega_{eff} \simeq 1$. As we can see, this solution for $\left|A\right|$ is the lower bound in the limit $\left|A\right| \geq 1/3$, which must be satisfied if we demand consistency with the SCM. Note that the dissipative effects are lost since the solution written in Eq. (\ref{eq:stiff}) do not depend on $\xi_{0}$, this can be corroborated if we calculate the viscous pressure, from Eq. (\ref{eq:devifrw}) one gets $\Pi(t) \rightarrow 0$, therefore, $p_{eff} \approx p$.  Note that in this regime the equilibrium condition $\left|\Pi / p \right| \ll 1$ is trivially satisfied. Using the fact that $\Pi(t) \rightarrow 0$ in Eq. (\ref{eq:dergen}), we can conclude the following   
\begin{equation}
\frac{dS}{dt} \rightarrow 0 \ \ \ \mbox{then} \ \ \ \frac{d^{2}S}{dt^{2}} = 0,
\end{equation}
yielding a positive and constant entropy given by
\begin{equation}
S \approx 6t_{s}\bar{\Gamma}\Theta,
\label{eq:entro}
\end{equation}
being $\bar{\Gamma}$ the re-definition of the constant $\Gamma$ (see Eq. (\ref{eq:entropy})) when we consider the solution $\left|A\right| \simeq 1/3$ and $\omega \approx 1$, besides $\Theta$ is also a constant. This can be seen from $\Theta = (t/t_{s}-1)^{\alpha}$, where the exponent $\alpha$ will be a very small and negative number near the stiff matter regime, then as time increases $\Theta \approx 1$. In summary, the viscous pressure, $\Pi$, tends to zero therefore the bulk viscosity effects are not dominant in this regime, additionally, as time increases the entropy tends to a constant and positive value.

\section{Final remarks} 
\label{sec:elfinal}
In this work we present a new solution of the full IS theory for an universe described by a single fluid which can have two different kind of cosmic expansion: accelerated, for $\left|A\right| > 1$, or decelerated, for $ 1/3 \leq \left|A\right| < 1$. One important feature of this solution is that the viscosity drives the effective EoS to be of quintessence type, for the single fluid with positive pressure. The solutions represent scale factors that take a zero value in the past, with a divergent curvature scalar, representing a Big Bang singularity.\\ 

In this depiction was found that cosmic expansion has a beginning and it is always due to an initial singularity. After inserting the new Ansatz in the transport equation of the IS theory we obtain a quadratic algebraic equation for $\left|A\right|$, then two branches were explored, by considering the thermodynamical criteria of growth of the entropy ($dS/dt > 0$) and the convexity condition ($d^{2}S/dt^{2} < 0$) was found that only one branch of the solution satisfies both conditions. An interesting characteristic of this branch is that it contains the three different types of cosmic expansion and each of these cases can be obtained by considering specific values ​​for the parameters of the model.\\ 

For $s=1/2$ was found in Ref. \cite{CruzLepe} that a de Sitter scenario is not admitted in the full IS theory. On the other hand, once we consider this new solution in the full IS framework we obtain that the late time behavior for this model is not exactly of a cosmological constant, such behavior can only be reached in the limit $\left|A\right|\rightarrow \infty$. By computing the curvature scalar, $R$, we obtain that in this viscous scheme has a positive sign for a decelerated universe or can turn negative for accelerated expansion, then, this change in the sign of $R$ could be used as a signal to identify the cosmic evolution we are dealing with, comparing with the standard cosmology such changes in the sign of the curvature scalar are not unusual. 
It is worthy to mention that in our approximation the analogous to an universe dominated by radiation can be found with one single value of our solution $\left|A\right|$, i.e., an universe with radiation content could be modelled using a viscous scheme. For other models with matter/viscous components as can be found in Refs. \cite{rad1, rad2}, the viscosity tends to be dominant during cosmic evolution and may even suppress the production of radiation.\\ 

At thermodynamical level we have a suitable behavior for this solution always that $1/3 \leq \left|A\right| < 1$, this was obtained for $s=1/2$ and using the general expression for the relaxation time, $\tau$. For an accelerated universe the convexity condition on entropy turns out to be positive, similar results were found at thermodynamical level in Ref. \cite{gong} for an universe with dark energy, however this was corrected by assuming a thermal equilibrium with the Hawking temperature between the universe and the apparent horizon, perhaps by implementing a method like this in the viscous scheme would help to clarify the thermodynamical behavior of the model, we will discuss this elsewhere.\\ 
It is important to mention that the bounds imposed by the thermodynamic criteria on the solution $\left|A\right|$, indicate that our approach is far from the equilibrium condition $\left|\Pi/p\right| \ll 1$, in both cases of accelerated and decelerated expansion. Then our solution shows that the fullfilment of the above condition can not be achieved even if the solution satisfies the thermodynamics conditions $dS/dt > 0$ and $d^{2}S/dt^{2} < 0$. In Ref. \cite{maartens} was proposed that for accelerated expansion one must take into account non-linearities in order to reach the aforementioned thermodynamic conditions. Our decelerated solution is indicating that these non-linearities must be also considered even when not accelerated expansion is present. This matter will be investigated in a future research.\\ 

It is well known that the evolution from a decelerated to an accelerated expansion (or vice versa) is characterized by a null deceleration parameter $q = 0$, in our formulation this value of $q$ is obtained when $\left|A\right| = 1$ and this represents a Milne universe where $\ddot{a} = 0$. Within the establishment of a decelerated universe we have an important case, $\left|A\right| = 1/2$, this value can be obtained only in the branch $\left|A\right|_{-}$ of the solution, for this spacetime we have for the curvature scalar, $R=0$, its scale factor evolution presents similarities with the evolution of a universe dominated by a radiation component in the SCM.\\ 

Finally, we explore the region near the stiff matter given by $\omega = 1$, in our case such value for $\omega$ can not be introduced formally since implicates singularities in some quantities of the IS theory such as the relaxation time for bulk viscous effects, but, as we approximate to this region we found that viscous effects dismiss, this is the only case where the equilibrium condition is satisfied, at thermodynamical level we obtain well behaved results.\\

\section*{Acknowledgments}
M. C. acknowledges the hospitality of the Instituto de F\'\i sica of Pontificia Universidad Cat\'olica de Valpara\'\i so, Chile, where part of this work was done. This work has been supported by S.N.I. (CONACyT-M\'exico) and PFCE-2016-UV (M.C.). This work has been supported by Fondecyt grant $\mbox{N}^{\circ}$ 1140238 (N. C.). \\
 
\appendix
\section{Stability of the solution}
\label{appest}
\begin{figure}[H]
\centering
\includegraphics[scale=0.45]{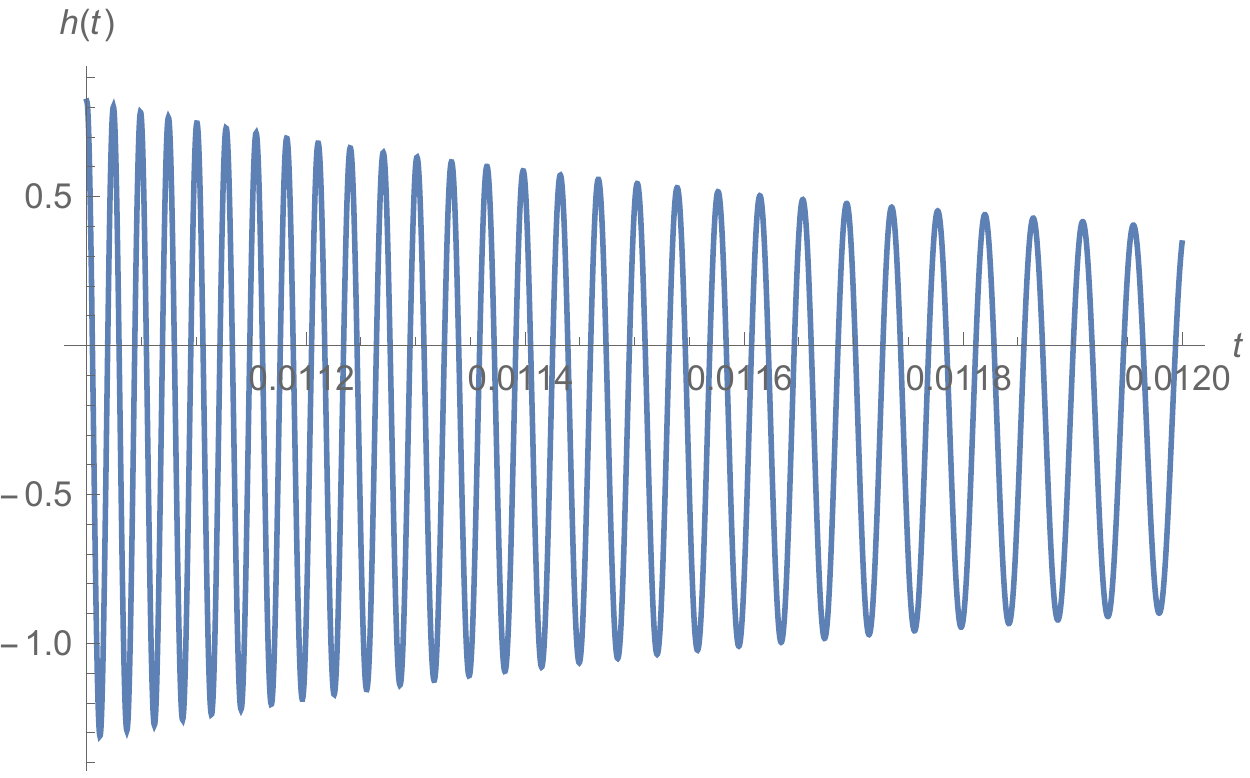}
\includegraphics[scale=0.45]{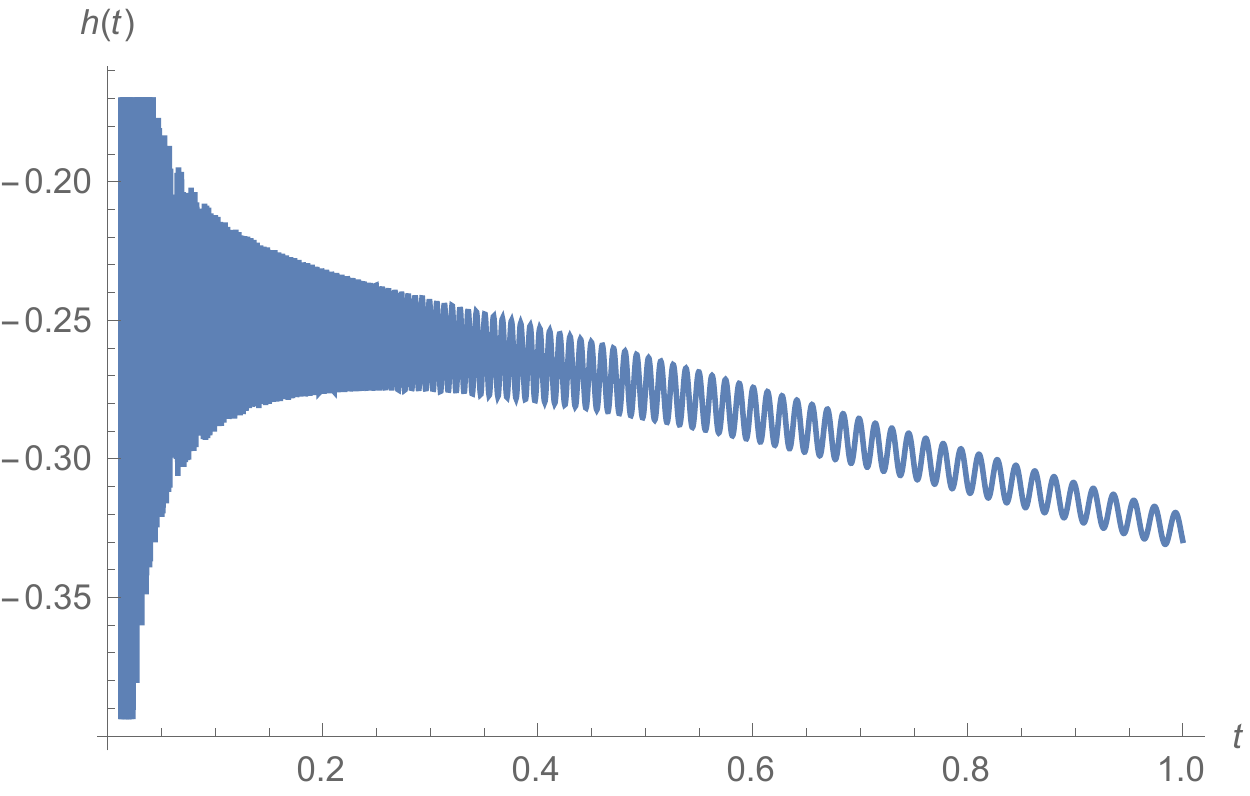}
\includegraphics[scale=0.45]{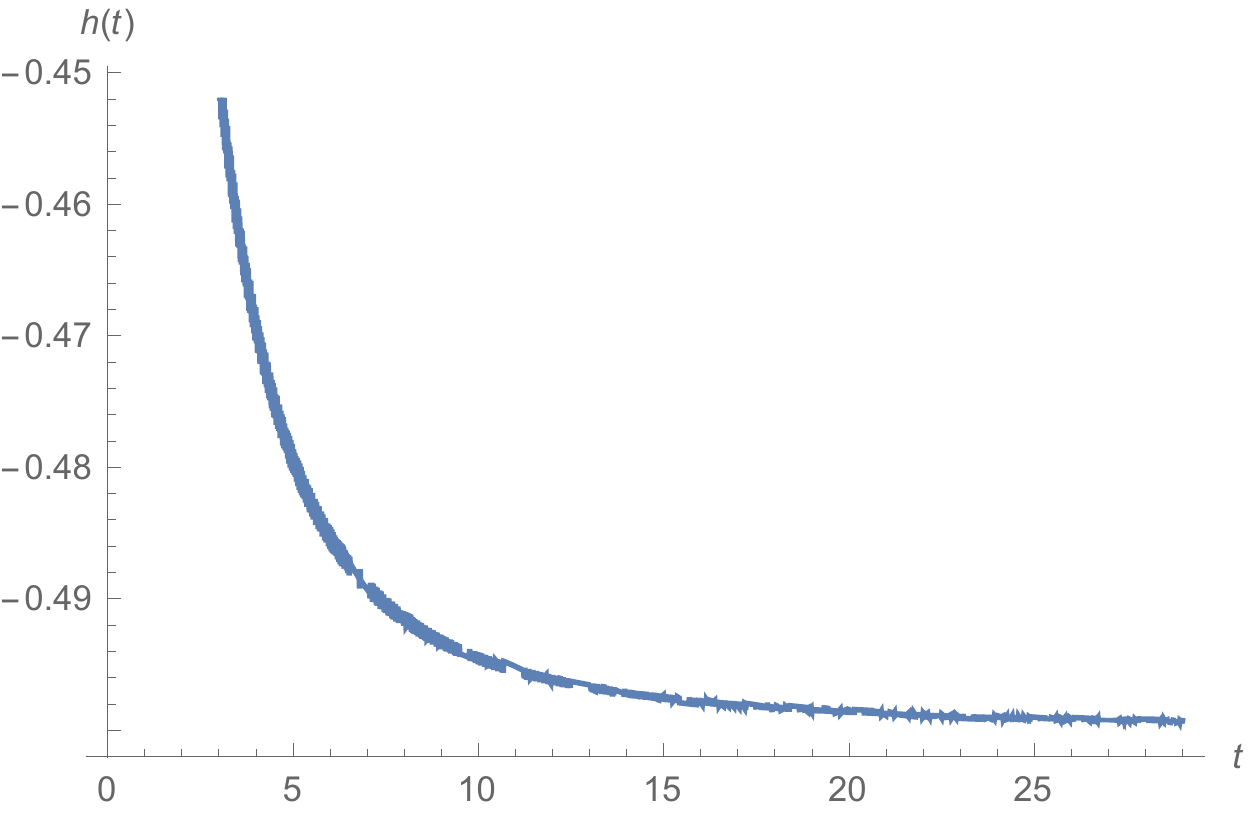}
\caption{Behavior of the perturbation $h(t)$ with $\left|A\right| < 1$.} 
\label{fig:pert1}
\end{figure}

In this appendix we discuss the stability of the solution written in Eq. (\ref{eq:quinte}) in a perturbatively way. We will consider the perturbation in the following form
\begin{equation}
H_{p}(t) = H(t)\left(1+h(t)\right), \ \ \ \mbox{where} \ \ \ \left|h\right| \ll 1,
\label{eq:pert}
\end{equation}
for consistency also the derivatives of the perturbation $h(t)$ are small. After inserting the perturbed solution given by Eq. (\ref{eq:pert}) in the transport equation (\ref{eq:is}) and considering the specific value $s=1/2$, we are left with the linearized differential equation for the perturbation, written as follows
\begin{small}
\begin{widetext}
\begin{align}
& \left[4\xi_{0}(1+\omega)(t-t_{s})^{2}\right]\ddot{h}(t)+4\xi_{0}(t-t_{s})\left[3\left|A\right|(1+\omega)+2\omega t_{s} \right]\dot{h}(t) - 4\left[9\left|A\right|\xi_{0}(1+\omega)t_{s} - 2\xi_{0}t^{2}_{s} -2\sqrt{3}\epsilon (1-\omega)(1+\omega)^{2}(t-t_{s})^{2}\right. \nonumber \\ 
& \left. - 3\left|A\right|^{2}(1+\omega)^{2}\left\lbrace 3\xi_{0}+2\epsilon (1-\omega)\left(\sqrt{3}(1+\omega)-3\xi_{0}\right)\right\rbrace\right]h(t) + 4\xi_{0}t^{2}_{s} - 12\left|A\right|\xi_{0}(1+\omega)t_{s} + 4\sqrt{3}\epsilon (1-\omega)(1+\omega)^{2}(t-t_{s})^{2}\nonumber \\
& + 3\left|A\right|^{2}(1+\omega)^{2}\left\lbrace 3\xi_{0}+2\epsilon (1-\omega)\left(\sqrt{3}(1+\omega)-3\xi_{0}\right)\right\rbrace = 0.
\label{eq:perturbed}
\end{align}
\end{widetext}
\end{small}
We defined $\dot{f} = df/dt$. The equation above can be solved numerically and we will restrict ourselves to the space of parameters $(\omega, \xi_{0}, \epsilon)$ where $\left|A\right| \in \Re$ as seen in Fig. (\ref{fig:absolution}). We will consider the value $10^{-4}$ for the parameter $\xi_{0}$. Since the solution $\left|A\right|_{-}$ can take values greater and less than 1, we will explore both regions, this can be done by varying the values of the parameter $\omega$. Additionally, we will consider a small value for $t_{s}$. In Fig. (\ref{fig:pert1}) by considering $\left|A\right| < 1$, we can observe a damped oscillatory behavior for the perturbation $h(t)$ and as times increase the solution keeps oscillating and tends to a minimum value. The maximum value of the the amplitude only depends on the initial conditions given for $h_{t=t_{0}}$ and $\dot{h}_{t=t_{0}}$, being $t_{0}$ the initial time and we have considered $t_{0} > t_{s}$. 
\onecolumngrid

\begin{figure}[H]
\centering
\includegraphics[scale=0.45]{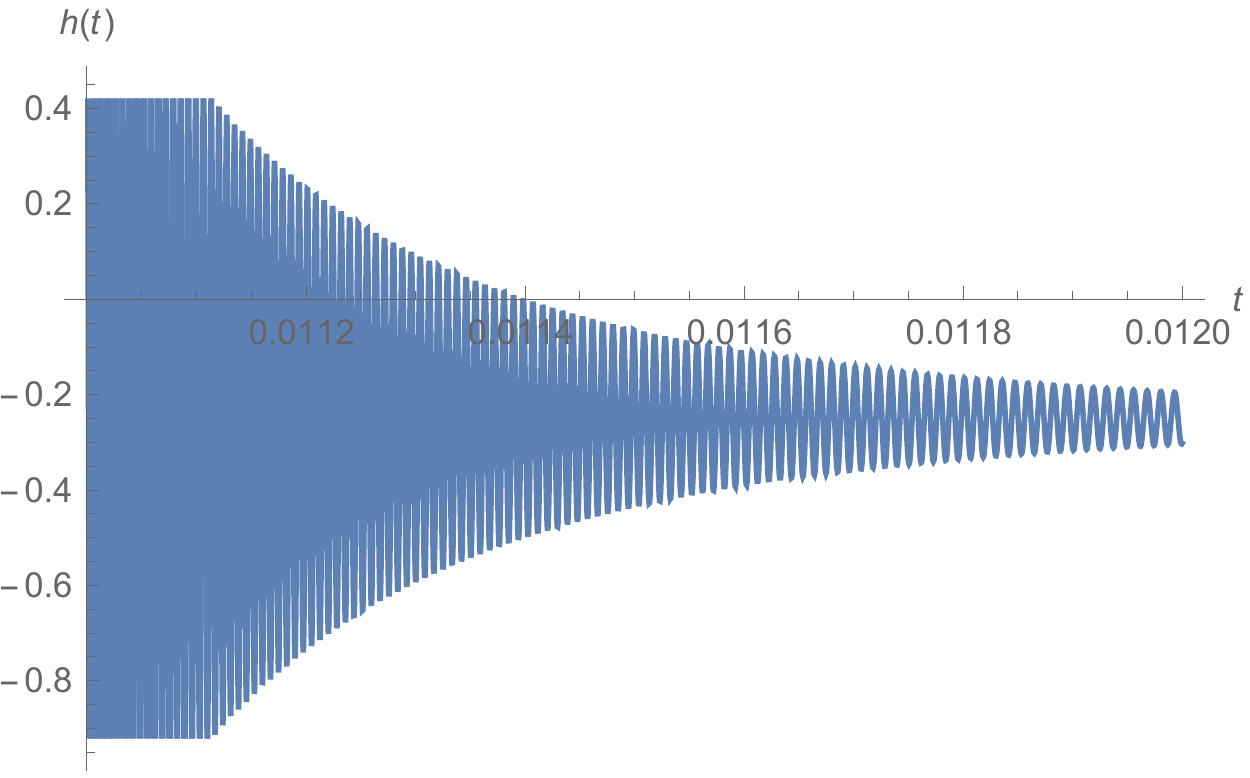}
\includegraphics[scale=0.45]{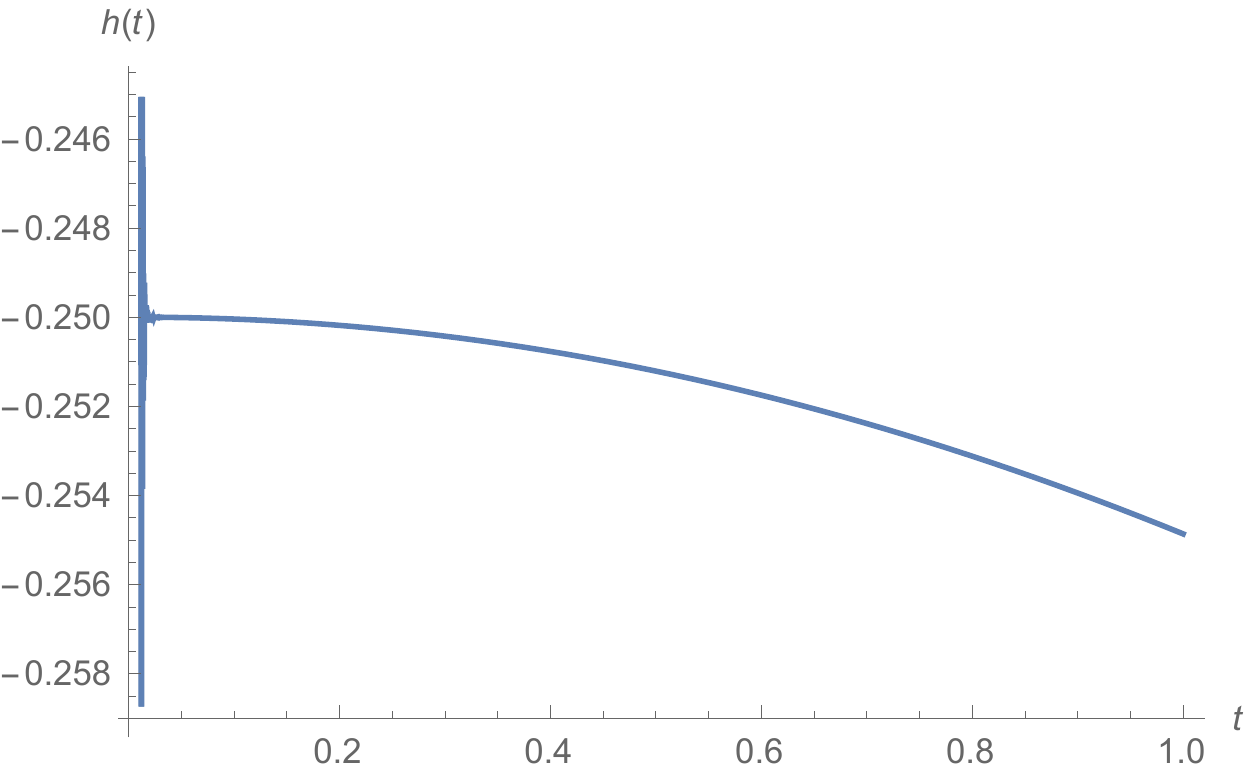}
\includegraphics[scale=0.45]{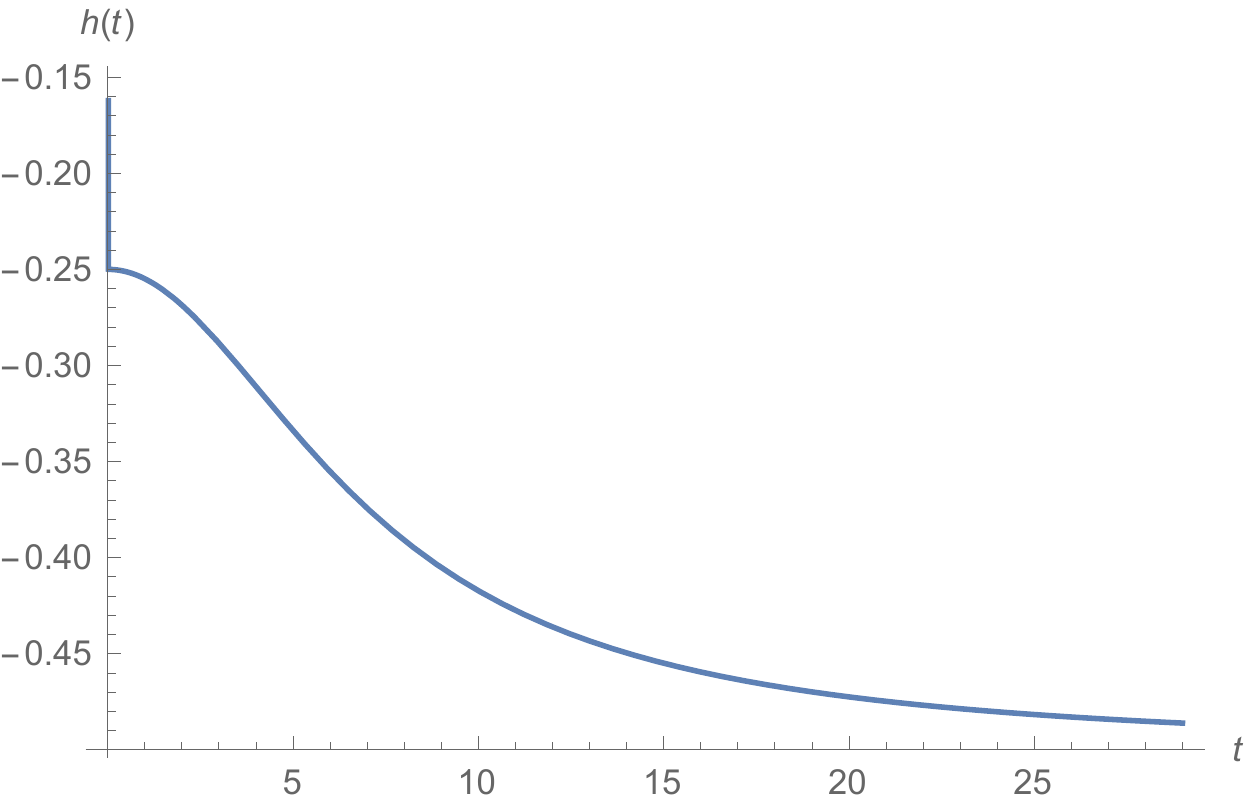}
\caption{Behavior of the perturbation $h(t)$ with $\left|A\right| > 1$.} 
\label{fig:pert2}
\end{figure}
\twocolumngrid
In Fig. (\ref{fig:pert2}) we depict the behavior of the perturbation with $\left|A\right| > 1$, for times around $t_{s}$ the solution presents a damped oscillatory behavior and rapidly tends to a minimum value (but do not oscillate around this minimum). In the figures we show the behavior of the perturbation with $\epsilon = 0.9$, for $\epsilon = 0.4, 0.2$ we obtain similar results for both cases.\\ 
On the other hand, if we consider the original space of parameters $(\omega, \xi_{0})$ but small values for the parameter $\epsilon$, we obtain that the perturbation has an increasing behavior for any time, therefore the perturbation it will present instabilities along its evolution, this unstable nature was obtained with $\epsilon = 10^{-4}$.

\end{document}